\documentclass{aa}

\usepackage{graphicx}
\usepackage{txfonts}
\usepackage[colorlinks=true, linkcolor=blue, citecolor=blue]{hyperref}
\usepackage{multirow}

\newcommand{\bdot}[1]{\overset{\scalebox{0.5}{$\bullet$}}{#1}}
\newcommand{\bbar}[1]{\overline{#1}}
\newcommand{\bast}{\scalebox{1.3}{$\,\ast\,$}}
\newcommand{\tidal}[1]{\left\{#1\right\}}
\newcommand{\real}[1]{\operatorname{Re}[#1]}
\newcommand{\imag}[1]{\operatorname{Im}[#1]}

\begin{document}

\title{Polar motion of Venus}
\author{P.-L.~Phan \and N.~Rambaux}
\institute{
    Sorbonne Université, Observatoire de Paris, Laboratoire-Temps-Espace, CNRS, France \\
    \email{pierre-louis.phan at obspm.fr}
}
\date{~}

\abstract
{Five Venus missions are under development to study the planet in the next decade, with both NASA's VERITAS and ESA's EnVision featuring a geophysical investigation among their objectives. Their radar and gravity experiments will determine Venus's orientation, enabling analyses of its spin dynamics to infer relevant geophysical and atmospheric properties.}
{This work aims to characterize Venus's polar motion, defined as the motion of its spin axis in a body-fixed frame. We focus on signatures from its interior and atmosphere to support potential detections of polar motion by future orbiters.}
{We developed a polar motion model for a triaxial planet accounting for solar torque, centrifugal and tidal deformations of a viscoelastic mantle, and atmospheric dynamics. Core-mantle coupling effects were analyzed separately, considering a simplified spherical core. We computed the period and damping time of the free motion (i.e., the Chandler wobble) and determined the frequencies and amplitudes of the forced motion.}
{We revisited the Chandler frequency expression. Solar torque is the dominant phenomenon affecting Venus's Chandler frequency, increasing it by a factor of 2.75, whereas solid deformations decrease it by less than 1.5\%. Our model predicts a Chandler period in the range $[12\,900 ; 18\,800]$ years (core not fully crystallized) or $[18\,100 ; 18\,900]$ years (core fully crystallized). The Chandler wobble appears as a linear polar drift of about 90 meters on Venus's surface during EnVision's four-year primary mission, at the limit of its resolution. We also predict forced polar motion oscillations with an amplitude of about 20 meters, driven by the atmosphere and the solar torque.}
{Compared to the 240 meter spin axis precession occurring in inertial space over this duration, these results suggest that Venus's polar motion could also be detectable by future orbiters. Polar motion should be incorporated into rotation models when anticipating these missions, providing additional constraints on the interior structure of Venus.\vspace{8.3mm}}

\maketitle

\section{Introduction}

\par Knowledge of the interior structure of Venus is needed to understand its history and to gather insights into why its evolutionary path diverged so significantly from that of Earth. Magellan, the most recent mission to Venus with a specific focus on its interior, enabled measurements of its gravity field \citep{konopliv1999venus} and tidal deformations \citep{konopliv1996venusian}, but lacked the precision needed to determine the state of the core \citep{dumoulin2017tidal}. Current plausible interior models for Venus consider a core that can be entirely liquid, entirely solid, or composed of a liquid outer core with a solid inner core \citep{dumoulin2017tidal, shah2022chemical, saliby2023viscosity}.

\par Examining an object's rotational dynamics can offer clues about the nature and properties of its interior. The motion of a planet's spin axis can be described relative to the fixed celestial sphere (precession and nutation) or relative to the planetary surface itself (polar motion). Both motions bring complementary constraints for interior models. Venus's spin axis precession was first measured by \citet{margot2021spin} using ground-based observations, providing an estimate of its moment of inertia which imposed loose constraints on the size of the core. This is a powerful technique, as it measures the orientation of a planet's spin axis in inertial space without requiring direct observation of the planetary surface. However, as a consequence, it did not measure the orientation of the spin pole relative to the surface.

\par The free component of a planet's polar motion is directly linked to its internal properties. On Earth, it was first measured by \citet{chandler1891variation} and has since been named the "Chandler wobble". Among the other planets, this free motion has only been detected on Mars \citep{konopliv2020detection}, which allowed to constrain the mantle rheology at a larger period than other tidal periods.
Polar motions of the proto-planets (1)~Ceres and (2)~Vesta have been computed in perspective of the Dawn mission, but were not detected \citep{rambaux2011constraining, rambaux2013rotational}.
During the early Pioneer Venus Orbiter mission, \citet{yoder1979venus} studied the hypothesized Chandler wobble of Venus, focusing on its amplitude (i.e., the angle between the spin axis and the polar principal axis of inertia), but that angle could not be measured. Since a non-zero amplitude is not a stable state, they explored the potential value of this amplitude by characterizing the wobble excitation and estimating its damping efficiency. Based on the degree-2 gravity field obtained later by the Magellan mission, this angle was measured at $0.5\degr$ \citep{konopliv1999venus}. The mechanisms that could excite Venus's polar motion to that amplitude are still not fully understood, although mantle convection has been studied as a potential excitation source \citep[e.g.,][]{spada1996spin}. Measuring the period of Venus's Chandler wobble, which offers a direct link to the planet's interior, requires detecting the motion of its spin pole, rather than solely measuring its position.

\par With a spacecraft in orbit around Venus, signals of its precession and polar motion could be detected by tracking surface radar features, or through the gravitational effect of spin dynamics on the spacecraft trajectory. The Magellan mission lacked the precision to detect these signals with either approach, but the detection of the precession has already been simulated for upcoming missions \citep{rosenblatt2021determination, cascioli2021determination}. However, detecting Venus's polar motion has not yet been addressed with similar analyses.

\par As a step toward this goal, we present a complete model for Venus's polar motion, by determining its relationship with the planet's interior and its atmospheric dynamics. In this work, Venus is modeled with a spherical core, a solid deformable triaxial mantle, and a fluid atmosphere. In Sect.~\ref{sec:model}, we describe the framework used to study the dynamics of the spin axis of Venus's mantle in a body-fixed reference frame: the Euler-Liouville equation, which forms the basis of our derivation of polar motion, and the relevant physical phenomena; namely, the solar torque, solid deformations, and atmospheric dynamics. Their individual effects on polar motion are presented in Sect.~\ref{sec:results}, where we characterize the Chandler period and damping time, and the forced polar motion, as applied to Venus. In Sect.~\ref{ssec:core}, we discuss the impact of different assumptions for core-mantle dynamical coupling and how they relate to the physical state of the core. Then, applications of the polar motion model to Venus, Earth and Mars are compared in Sect.~\ref{ssec:earth-mars}. Finally, in Sect.~\ref{ssec:envision}, we discuss the potential for future missions, focusing on EnVision, to detect and measure Venus's polar motion. We also discuss the constraints this could provide on Venus's internal structure.

\section{Polar motion model}
\label{sec:model}

\begin{figure*}
    \centering
    \includegraphics[width=0.495\hsize]{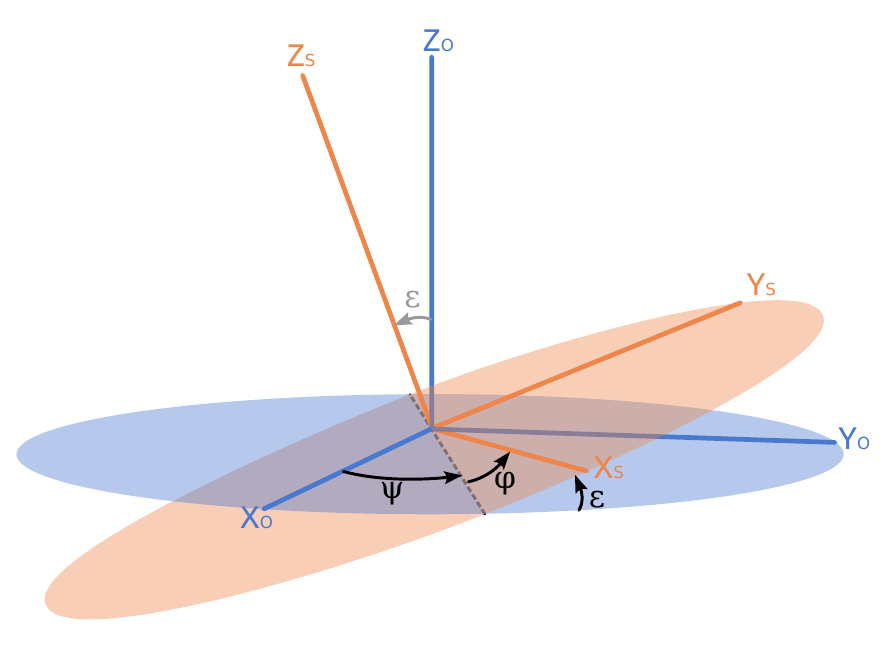}
    \includegraphics[width=0.495\hsize]{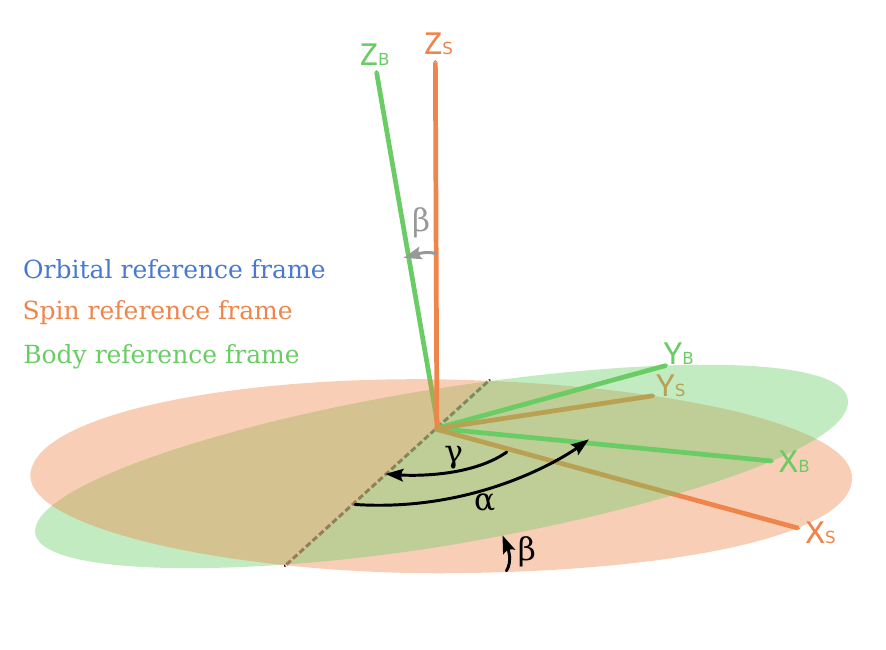}
    \caption{Transformations between the ORF (blue), SRF (orange), and BRF (green), with Euler angles. The SRF and the BRF are both rotating frames, with $Z_S$ the spin axis of Venus and $Z_B$ its polar principal axis. The polar motion is the relative motion between the SRF and the BRF.}
    \label{fig:frames}
\end{figure*}

\subsection{Linearized Euler-Liouville equation}

\begin{table*}
    \caption{Composite angles involved in the forcing of Venus's polar motion and associated Love numbers.} \label{tab:frequencies}
    \centering \renewcommand{\arraystretch}{1.4}
    \begin{tabular}{l r r c r r}
    \hline \hline \noalign{\smallskip}
        Angle & Frequency & Period [days] & Tidal coefficient & $\lvert k_2^{(j)} \rvert$ & $\delta_j ~{[\degr]}$ \\
    \noalign{\smallskip} \hline \noalign{\smallskip}
        $\theta_1 = -\phi - \gamma - \alpha$ & $-\Omega$ & $243.0$ & $k_2^{(1)}$ & $0.303$ & $-1.9$ \\
        $\theta_2 = 2(M - \psi) - \phi - \gamma - \alpha$ & $2n-\Omega$ & $76.8$ & $k_2^{(2)}$ & $0.298$ & $-1.4$ \\
        $\theta_3 = 2(M - \psi - \phi - \gamma) - \alpha$ & $2(n-\Omega)$ & $58.4$ & $k_2^{(3)}$ & $0.296$ & $-1.3$ \\
        $-\alpha$ & $\sigma_\mathrm{cw}$ & $\approx -6 \times 10^6$ & $k_2^{(\mathrm{cw})}$ & $[0.48~;~0.93]$ & $[0.18~;~24]$ \\
    \noalign{\smallskip} \hline
    \end{tabular}
    \tablefoot{The fast angles are $M$ and $\phi$. Indicative values of $k_2$ and phase lags $\delta$ (Y. Musseau, priv. comm. 2024) are also listed.}
\end{table*}

\par We analyzed the polar motion of Venus, defined as the motion of its spin axis in a body-fixed frame, via the Euler rotation equation. It describes the dynamics of angular momentum $\vec{H}$ in a non-inertial reference frame with
\begin{equation} \label{eq:euler}
    \vec{\bdot{H}} + \vec{\omega} \times \vec{H} = \vec{\Gamma} \,,
\end{equation}
where $\vec{\omega}$ is the instantaneous rotation vector of that reference frame and $\vec{\Gamma}$ is the external torque. The polar motion is obtained by solving this equation for $\vec{\omega}$ in a Venus body-fixed frame, such that $\vec{\omega}$ also describes the rotation of the planet \citep{munk1960rotation, gross2015earth, rambaux2013rotational}.

\par We worked in the body-fixed frame in which the time-averaged inertia matrix, $\mathcal{I}_0$, is diagonal, referred to as the body reference frame (BRF), whose axes are the principal axes of inertia of the planet. In this reference frame, Venus's spin axis remains close to the z-axis \citep[with a present-day angle of $0.481\degr \pm 0.020\degr$,][]{konopliv1999venus}, so the rotation vector can then be written as variations, $\vec{\Delta\omega}$, to the main z-axis component, $\vec{\omega}_0$:
\begin{equation} \label{eq:omega}
    \vec{\omega} = \vec{\omega_0} + \vec{\Delta\omega} = \begin{pmatrix} 0 \\ 0 \\ \Omega \end{pmatrix} + \Omega \begin{pmatrix} m_x \\ m_y \\ m_z \end{pmatrix} \,,
\end{equation}
where $\Omega$ is the constant mean rotation rate of Venus, and $m_x$, $m_y$, and $m_z$ are treated as small quantities. In this work, the retrograde rotation of Venus appears as a negative $\Omega$, with an obliquity that is between $0\degr$ and $90\degr$ \citep[$\epsilon = 2.6392\degr \pm 0.0008\degr$,][]{margot2021spin}, but switching to the opposite convention does not affect the results. The values of parameters used for Venus in this work are listed in Table~\ref{tab:venus_numbers}.

\par To solve Eq.~(\ref{eq:euler}) for $\vec{\omega}$ we express the angular momentum as
\begin{equation}
     \vec{H} = \mathcal{I} \vec{\omega} + \vec{h} \,,
\end{equation}
where $\mathcal{I}$ is the inertia matrix of the planet including the atmosphere, so that $\mathcal{I} \vec{\omega}$ represents the angular momentum of a uniformly rotating planet including an atmosphere at rest. The winds contribute additional angular momentum, $\vec{h}$. When written in the BRF, the inertia matrix, $\mathcal{I}$, is decomposed into a mean term, $\mathcal{I}_0$, and the small deformation term $\Delta\mathcal{I}$ originating from both mantle and atmospheric deformation, with
\begin{equation} \label{eq:inertia}
    \mathcal{I} = \mathcal{I}_0 + \Delta\mathcal{I}
    = \begin{bmatrix} A & 0 & 0 \\ 0 & B & 0 \\ 0 & 0 & C \end{bmatrix}
    + \begin{bmatrix}
        \Delta\mathcal{I}_{xx} & \Delta\mathcal{I}_{xy} & \Delta\mathcal{I}_{xz} \\
        \Delta\mathcal{I}_{xy} & \Delta\mathcal{I}_{yy} & \Delta\mathcal{I}_{yz} \\
        \Delta\mathcal{I}_{xz} & \Delta\mathcal{I}_{yz} & \Delta\mathcal{I}_{zz}
        \end{bmatrix} \,.
\end{equation}
In this work, the mean moments of inertia $(A, B, C)$ are either those of the whole planet (core, mantle and atmosphere) if the core is assumed to be dynamically locked with the mantle; or those without the core, if it is decoupled from the mantle (see Sect.~\ref{ssec:core}).

\par As is the case for Venus, the atmosphere is in a state of global super-rotation when the excess angular momentum $\vec{h}$ from the winds has a non-zero average. Here, this excess angular momentum is assumed to have constant amplitude and to take place around the planet's spin axis. It is noted $\vec{h}_{sr}$, and the small variations from atmospheric dynamics are denoted as $\vec{\Delta h}$:
\begin{equation} \label{eq:relative_momentum}
    \vec{h} = \vec{h}_{sr} + \vec{\Delta h} = k\vec{\omega} + \begin{pmatrix} \Delta h_x \\ \Delta h_y \\ \Delta h_z \end{pmatrix} \,,
\end{equation}
where $k$ is the scalar coefficient between $\vec{h}_{sr}$ and $\vec{\omega}$. In the absence of super-rotation, $k = 0$. For Venus, $k/C \approx 10^{-3}$ (see Sect.~\ref{ssec:superrotation}). With a first-order expansion in small quantities, the effects of atmosphere super-rotation are too minor to be taken into account here. Still, in Sect.~\ref{ssec:superrotation} we retain the terms with $k$ to discuss how super-rotation affects polar motion.

\par Using Eqs.~(\ref{eq:omega}-\ref{eq:inertia}) in Eq.~(\ref{eq:euler}), and keeping only the first-order terms in $\vec{\Delta\omega}$, $\Delta\mathcal{I}$ and $\vec{h}$, we have the Euler-Liouville equations \citep[e.g.,][chap.~6]{munk1960rotation}:
\begin{equation} \label{eq:liouville_xyz} \begin{aligned}
     A \bdot{m_x} + \bdot{\Delta\mathcal{I}_{xz}} + \frac{1}{\Omega} \bdot{\Delta h_x} + \Omega (C\!-\!B) m_y - \Omega \Delta\mathcal{I}_{yz} - \Delta h_y & = \frac{1}{\Omega} \Gamma_x \\
     B \bdot{m_y} + \bdot{\Delta\mathcal{I}_{yz}} + \frac{1}{\Omega} \bdot{\Delta h_y} - \Omega (C\!-\!A) m_x + \Omega \Delta\mathcal{I}_{xz} + \Delta h_x & = \frac{1}{\Omega} \Gamma_y \\
     C \bdot{m_z} + \bdot{\Delta\mathcal{I}_{zz}} + \frac{1}{\Omega} \bdot{\Delta h_z} & = \frac{1}{\Omega} \Gamma_z \,.
\end{aligned} \end{equation}
The third dimension is now decoupled from the others and can be studied separately to analyze $m_z$, the spin rate variations \citep{cottereau2011various}. The polar motion is described by $m_x$ and $m_y$, the projection of the unit rotation vector onto the equatorial plane of the BRF. The first two dimensions of Eq.~(\ref{eq:liouville_xyz}) are combined into a single complex equation by introducing the following complex notations: $m = m_x + i m_y$, $T = \Gamma_x + i \Gamma_y$, $\Delta I = \Delta\mathcal{I}_{xz} + i \Delta\mathcal{I}_{yz}$, and $\Delta h = \Delta h_x + i \Delta h_y$. This yields
\begin{equation} \label{eq:liouville} \begin{aligned}
     & \left[ \frac{A\!+\!B}{2} \bdot{m} - i \Omega \left(C - \frac{A\!+\!B}{2} \right) m \right] - \frac{B-A}{2} \left( \bdot{\bbar{m}} + i \Omega \bbar{m} \right) = \\
     & \hspace{9em} \frac{1}{\Omega} T - \left( i \Omega \Delta I + \bdot{\Delta I} \right) - \left( i \Delta h + \frac{1}{\Omega} \bdot{\Delta h} \right) \,,
\end{aligned} \end{equation}
where $\bbar{m}$ is the complex conjugate of $m$.

\par The Chandler wobble is the free oscillation that can be characterized by solving the homogeneous equation from Eq.~(\ref{eq:liouville}) for $m$, effectively ignoring the terms that are periodic with external forcing periods. However, some terms on the right-hand side of the equation depend on $m$, thus affecting the homogeneous system. They need to be identified for a complete description of the Chandler wobble. For instance, the solid rotational deformation term is proportional to $m$, lengthening the Earth's calculated wobble period by 38\% \citep[e.g.,][]{gross2015earth}. As pointed out by \citet[][Appendix 2]{baland2019coupling}, the solar torque term for a rigid planet also contains components proportional to $m$, but shortens the Earth's free wobble period by less than 0.001\%. In the sections below, we explore how these $m$-dependent terms affect Venus's free wobble.

\subsection{Solar torque influence on the rigid Chandler wobble}
\label{ssec:torque}

\par Here, we highlight how the solar torque affects the wobble by considering a simplified non-deformable solid body without relative atmospheric angular momentum, thus ignoring the terms in $\Delta I$ and $h$. Venus's orbit currently has an eccentricity of 0.0067, the lowest among Solar System planets; hence, our analysis assumes a circular orbit.

\par The solar torque exerted on Venus depends on the position $\vec{r}$ of the Sun and on Venus's inertia matrix, both expressed in the BRF. It can be written like in \citet{williams2001lunar} as
\begin{equation} \label{eq:torque_vec}
    \vec{\Gamma} = 3 \frac{GM_\sun}{\lVert\vec{r}\rVert^5} \vec{r} \times \mathcal{I} \vec{r} \,,
\end{equation}
where $G$ is the gravitational constant and $M_\sun$ the mass of the Sun. By neglecting the deformations and the eccentricity, we find the complex representation of the torque:
\begin{equation} \label{eq:torque}
    T_0 = 3 n^2 \left[ \left(C-\frac{A+B}{2}\right)(yz - i xz) - \frac{B-A}{2}(yz + i xz) \right] \,,
\end{equation}
with $n$ the mean orbital motion of Venus, and $\vec{r}/\lVert\vec{r}\rVert = (x, y, z)^T$ in the BRF. Then, to solve Eq.~(\ref{eq:liouville}), we need to derive analytical expressions for the products $yz$ and $xz$.

\par We define an orbital reference frame (ORF) centered on Venus. Its xy-plane coincides with Venus's orbital plane, and the ascending node of this plane on the ICRF equator defines the ORF x-axis. In this work, the effects of other planets on Venus's orbit are neglected, allowing the ORF to be considered an inertial reference frame. For a circular orbit, with $M = nt + M_0$ the mean anomaly of Venus, the planet's coordinates in its own orbital plane are given by $\lVert\vec{r}\rVert(\cos M,\, \sin M,\, 0)^T = \mathcal{R}_z(-M)(\lVert\vec{r}\rVert, 0, 0)^T$, using the elementary rotation matrices:
\begin{equation}
    \mathcal{R}_z(\theta) = \begin{bmatrix} \cos\theta &\!\! \sin\theta &\!\! 0 \\ -\sin\theta &\!\! \cos\theta &\!\! 0 \\ 0 &\!\! 0 &\!\! 1 \end{bmatrix} \text{ and }
    \mathcal{R}_x(\theta) = \begin{bmatrix} 1 &\!\! 0 &\!\! 0 \\ 0 &\!\! \cos\theta &\!\! \sin\theta \\ 0 &\!\! -\sin\theta &\!\! \cos\theta \end{bmatrix}.
\end{equation}
The coordinates of the Sun in the ORF, centered on Venus, are given by the opposite vector
\begin{equation} \label{eq:r-orf}
    \frac{\vec{r}_\mathrm{ORF}}{\lVert\vec{r}\rVert} = -\begin{pmatrix} \cos M \\ \sin M \\ 0 \end{pmatrix} = -\mathcal{R}_z(-M)\begin{pmatrix} 1 \\ 0 \\ 0 \end{pmatrix} \,.
\end{equation}

\par To obtain the coordinates of the Sun in the BRF, we use an intermediate rotating frame, referred to as the spin reference frame (SRF), whose z-axis coincides with the spin axis $\vec{\omega}$, and whose x-axis points toward Venus's prime meridian. The transformation from the ORF to the SRF is described with Euler angles, as illustrated in Fig.~\ref{fig:frames}:
\begin{equation}
    \vec{r}_\mathrm{SRF} = \mathcal{R}_z(\phi) \mathcal{R}_x(\epsilon) \mathcal{R}_z(\psi) \vec{r}_\mathrm{ORF} \,.
\end{equation}
Here, $\epsilon$ is the obliquity of Venus, $\phi = \Omega t + \phi_0$ is the rotation angle of its prime meridian, and $\psi$ is the longitude of the ascending node of Venus's equator on its orbital plane (i.e., the nodal precession angle).

\par Due to the non-zero polar motion, the SRF is not a body-fixed reference frame. When the inertia matrix of Venus is expressed in the SRF, its eigenvectors make up the rows of the matrix transformation from the SRF to the BRF, which we describe with Euler angles $(\alpha, \beta, \gamma)$:
\begin{equation} \label{eq:srf-brf}
    \vec{r}_\mathrm{BRF} = \mathcal{R}_z(\alpha) \mathcal{R}_x(\beta) \mathcal{R}_z(\gamma) \vec{r}_\mathrm{SRF} \,.
\end{equation}
They can be derived from a set of degree-2 gravity coefficients that uses the SRF, as in \citet{konopliv1999venus}, and the values of $(\alpha, \beta, \gamma)$ for Venus at epoch J2000 are given in Table~\ref{tab:venus_numbers}. Here, $\beta$ is the angle between the spin axis and the polar principal axis of inertia (i.e., the amplitude of polar motion). Both angles $\alpha$ and $\gamma$ evolve with the phase of polar motion, with $\alpha$ increasing and $\gamma$ decreasing in the case of Venus. The configuration depicted in Fig.~\ref{fig:frames} is similar to that at epoch J2000, where $\alpha_{J2000}$ is positive and $\gamma_{J2000}$ is negative. The sum $\alpha + \gamma$ represents the longitudinal offset between the smallest principal axis of inertia of Venus and its prime meridian, measured at $-3.17\degr$ \citep{konopliv1999venus}.

\par Using Eqs.~(\ref{eq:r-orf}-\ref{eq:srf-brf}), we have a complete representation of the coordinates of the Sun in the BRF:
\begin{align}
    \frac{\vec{r}_{BRF}}{\lVert\vec{r}\rVert} & = -\mathcal{R}_z(\alpha) \mathcal{R}_x(\beta) \mathcal{R}_z(\gamma) \hspace{0.5em} \mathcal{R}_z(\phi) \mathcal{R}_x(\epsilon) \mathcal{R}_z(\psi) \hspace{0.5em} \mathcal{R}_z(-M) \begin{pmatrix} 1 \\ 0 \\ 0 \end{pmatrix} \\
    \frac{\vec{r}_{BRF}}{\lVert\vec{r}\rVert} & = \begin{pmatrix} x \\ y \\ z \end{pmatrix} = \begin{pmatrix} -\cos\Big( (M - \phi) - (\psi + \gamma + \alpha) \Big) \\ -\sin\Big( (M - \phi) - (\psi + \gamma + \alpha) \Big) \\ \epsilon \sin(M - \psi) + \beta \sin\Big( (M - \phi) - (\psi + \gamma) \Big) \end{pmatrix} \label{eq:sun_xyz} \,,
\end{align}
where we considered only the first-order terms in $\epsilon$ and $\beta$. Here $M$ and $\phi$ are the fast-rotating angles, with respectively a 224.7 days orbital period and a -243.0 days rotation period. The other angles are much slower, namely, the precession angle $\psi$ has a $\approx 29\,000$ years period \citep{margot2021spin}, and both $-\alpha$ and $\gamma$ have periods corresponding to the Chandler period which we find to be $\approx 16\,000$ years. The products $xz$ and $yz$ are found to contain terms with the three fast angles $\theta_j$ defined in Table~\ref{tab:frequencies}, but also contain a term with the angle $\alpha$:
\begin{equation} \label{eq:products} \begin{aligned}
    xz & = \frac{\epsilon}{2} (\sin\theta_1 - \sin\theta_2) - \frac{\beta}{2} (\sin\theta_3 + \sin\alpha) \\
    yz & = - \frac{\epsilon}{2} (\cos\theta_1 - \cos\theta_2) + \frac{\beta}{2} (\cos\theta_3 - \cos\alpha) \\
    xy & = \frac{1}{2} \sin(\theta_3 - \alpha) \,.
\end{aligned} \end{equation}
Moreover, from the definition of both reference frames we have: $\vec{\omega}_\mathrm{BRF} = \Omega (m_x, m_y, 1+m_z)^T$ and $\vec{\omega}_\mathrm{SRF} = \Omega (0, 0, 1)^T$. Hence, the Euler angles $\alpha$ and $\beta$, involved in the transformation Eq.~(\ref{eq:srf-brf}) describing the polar motion, are linked to $m_x$ and $m_y$ with
\begin{equation}
    m = m_x + i m_y = \beta (\sin\alpha + i \cos\alpha) = i \beta e^{-i\alpha} \,.
\end{equation}
The torque Eq.~(\ref{eq:torque}) is thus found to be
\begin{equation} \label{eq:T0} \begin{aligned}
    T_0 & = i \frac{3}{2} n^2 \left[ \left(C-\frac{A+B}{2}\right) m + \frac{B-A}{2} \bbar{m} \right] \\
        & \hspace{4em} + \frac{3}{2} n^2 \left(C-\frac{A+B}{2}\right) \left[ \epsilon \left( - e^{i\theta_1} + e^{i\theta_2} \right) + \beta e^{i\theta_3} \right] \\
        & \hspace{4em} + \frac{3}{2} n^2 \frac{B-A}{2} \left[ \epsilon \left( e^{-i\theta_1} - e^{-i\theta_2} \right) - \beta e^{-i\theta_3} \right] \,.
\end{aligned} \end{equation}
The components in $\theta_j$, in the last two lines, contain the fast angles. Thus, they are unrelated to the Chandler wobble and treated as forcing terms, as described in Sect.~\ref{ssec:forced}.

\par On the other hand, the terms in the first line are proportional to $m$ and $\bbar{m}$ ; we substitute these terms in the Euler equation (\ref{eq:liouville}):
\begin{align}
     & \left[ \frac{A+B}{2} \bdot{m} - i \Omega \left(C - \frac{A+B}{2} \right) m \right] - \frac{B-A}{2} \left( \bdot{\bbar{m}} + i \Omega \bbar{m} \right) \\
     & \hspace{9em} - i \frac{3}{2} \frac{n^2}{\Omega} \left[ \left(C-\frac{A+B}{2}\right) m + \frac{B-A}{2} \bbar{m} \right] = 0 \nonumber \,.
\end{align}
It lets us find the free rigid Chandler wobble solution:
\begin{equation} \label{eq:chandler_motion_rigid}
    m(t) = m_0 \exp(i \sigma_r \, t) - \frac{\kappa-1}{\kappa+1} \bbar{m_0} \exp(-i \sigma_r \, t) \,,
\end{equation}
which can be written as
\begin{equation} \begin{aligned}
    m_x(t) & = m_{x0} \cos(\sigma_r \, t + \eta) \\
    m_y(t) & = \kappa \, m_{x0} \sin(\sigma_r \, t + \eta) \,,
\end{aligned} \end{equation}
with the frequency and the ellipticity of the wobble being
\begin{align} \label{eq:chandler_freq_rigid}
    \sigma_r & = \Omega \sqrt{\frac{C-A}{A} \, \frac{C-B}{B}} \left( 1 + \frac{3}{2} \frac{n^2}{\Omega^2} \right) \,, \\
    \kappa & = \sqrt{\frac{C-A}{C-B} \, \frac{A}{B}} \,,
\end{align}
and where the complex initial condition $m_0$, or the real ones $(m_{x0}, \eta)$, must be constrained by observations. The triaxiality of the planet results in an elliptic motion as shown in Fig.~\ref{fig:chandler_wobble}, with $\kappa > 1$, and with the presence of the negative Chandler frequency in Eq.~(\ref{eq:chandler_motion_rigid}). As previously discussed, the values to use for $(A, B, C)$ should either be those of the whole planet (core, mantle and atmosphere) or those without the core, depending on the core-mantle coupling assumption.

\par Equation~(\ref{eq:chandler_freq_rigid}) for the Chandler frequency accounts for the solar torque, aligning with the expressions already obtained by \citet{yoder1997venusian} and \citet[][Appendix 2]{baland2019coupling}, while also accounting for triaxiality as done in \citet{rambaux2013rotational}. In Sect.~\ref{ssec:deformation}, we extend it to incorporate the effects of solid deformation. The key point here is that the solar torque, usually ignored when studying the free Chandler wobble, actually increases the Chandler frequency by a factor of $1 + (3n^2)/(2\Omega^2)$, which is about 2.75 when applied to Venus. For Earth and Mars, which rotate faster, this factor has respective values of $(1+10^{-5})$ and $(1+3\times10^{-6})$ and can be considered negligible. Another way to understand the contribution of the solar torque is to develop the Chandler frequency Eq.~(\ref{eq:chandler_freq_rigid}) as
\begin{equation} \label{eq:chandler_freq_precession}
    \sigma_r = \Omega \sqrt{\frac{C-A}{A} \, \frac{C-B}{B}} + q \frac{3}{2} \frac{n^2}{\Omega} \frac{C - \frac{A+B}{2}}{C} \,,
\end{equation}
where $q \!=\! \frac{C}{\sqrt{AB}} \frac{2\sqrt{(C-A)(C-B)}}{(C-A) + (C-B)}$ accounts for the non-sphericity of the planet, and is close to $1$. The first term is the usual formulation of the Chandler frequency which ignores the solar torque \citep[e.g.,][]{rambaux2013rotational, konopliv2020detection}. Neglecting triaxiality ($q = 1$) and assuming a small obliquity, we recognize in the second term the frequency of the precession due to solar torque \citep[e.g.,][]{williams1994contributions}. Comparing the two terms of Eq.~(\ref{eq:chandler_freq_precession}), in the cases of Earth and Mars, the torque-free wobble has a relatively high frequency (on the order of a few hundred days) compared to that of the solar-induced precession, which is on the order of a hundred thousand years, making the second term negligible. In the case of Venus, the torque-free wobble and solar-induced precession have periods of the same order, respectively, $\approx 51\,000$ years and $\approx 29\,000$ years, so both terms in Eq.~(\ref{eq:chandler_freq_precession}) should be kept, resulting in a Chandler wobble period shorter than both.

\par The expression in Eq.~(\ref{eq:chandler_freq_rigid}) for the Chandler frequency, $\sigma_r$, is valid for a non-deformable solid body, where ignoring the dissipative processes leads to the absence of damping for the Chandler wobble. In the next section, we introduce solid deformation into the model.

\subsection{Solid deformations}
\label{ssec:deformation}

\par The solid viscoelastic deformation of Venus's solid body affects polar motion in two distinct ways. A direct effect is due to angular momentum conservation and is seen in Eq.~(\ref{eq:liouville}) with the terms in $\Delta I$ and its derivative; and indirectly, variations of $\mathcal{I}$ can induce variations in the net solar torque, as shown in Eq.~(\ref{eq:torque_vec}). Here, we explore two sources of solid deformation for Venus: the centrifugal potential and the solar tidal potential, which cause variations $\Delta\mathcal{I}_R$ and $\Delta\mathcal{I}_T$ of the inertia matrix. They result in additional torque terms, respectively $\vec{\Gamma}_R$ and $\vec{\Gamma}_T$. All of these terms are taken into account in Eq.~(\ref{eq:liouville}) with $\Delta I = \Delta I_R + \Delta I_T$ and $T = T_0 + T_R + T_T$\,.

\subsubsection{Love numbers formalism}
\par The tidal response $\Delta\mathcal{I}$ of a body to a perturbation $\Delta\mathcal{I}^p$ depends on the history of said perturbation: $\Delta\mathcal{I}(t) = \tilde{k_2}(\tau) \bast \Delta\mathcal{I}^p(t)$, where the symbol $\bast$ designates the temporal convolution, and $\tilde{k_2}(\tau)$ is a convolution kernel that depends on the rheology of the body \citep[e.g.,][]{boue2016complete, correia2022tidal}.

\par In practice, when the excitation is decomposed as a set of distinct periodic components, it means that components of different frequencies $\nu_j$ are affected by different Love numbers $k_2^{(j)}$, which come from the Fourier transform of $\tilde{k_2}(\tau)$. These Love numbers have complex values $k_2^{(j)} \!=\! \lvert k_2^{(j)} \rvert \exp(i \delta_j)$, where the modules $\lvert k_2^{(j)} \rvert$ are the amplitude gains of Venus's response, and the arguments $\delta_j$ are the phase lags that characterize the inelastic response. Here, in equations where an excitation term contains multiple periodic components, we use the convolution notation for improved clarity. When the components are developed separately, we use the individual $k_2^{(j)}$ values.

\par To date, the tidal response of Venus has only been measured at the semi-diurnal period of 58 days \citep{konopliv1996venusian} and rheological models are needed to extrapolate this data to other frequencies. We use models provided by \citet{musseau2024viscosity}, where some have a constant mantle viscosity, ranging from $10^{19}$ Pa\,s to $10^{22}$ Pa\,s, and some have a viscosity profile which they fitted to match the current rotation rate of Venus, where the thermal atmospheric tides compensate the solid tides to an equilibrium. The $\lvert k_2^{(j)} \rvert$ and $\delta_j$ values from their four-layer viscosity profile and V5-T$_\mathrm{hot}$ interior model are given in Table~\ref{tab:frequencies}. Here, we use the convention that for positive frequencies $\delta_j$ is negative, whereas for negative frequencies, $\delta_j$ is positive.

\subsubsection{Centrifugal and tidal deformations}
\par We derive the centrifugal deformation $\Delta\mathcal{I}_R$ caused by Venus's rotation and the tidal deformation $\Delta\mathcal{I}_T$ caused by the gravity of the Sun. We followed the expressions of \citet{williams2001lunar} where we removed the average components, already contained in $\mathcal{I}_0$, and retained only the first-order terms in $\beta$ and $\epsilon$. It leads to the following expressions:
\begin{equation} \label{eq:ir_it_mat} \begin{aligned}
    \Delta\mathcal{I}_R & = \tilde{k_2} \bast \frac{\Omega^2 R^5}{3G} \begin{bmatrix} -\frac{2}{3}m_z & 0 & m_x \\ 0 & -\frac{2}{3}m_z & m_y \\ m_x & m_y & \frac{4}{3}m_z \end{bmatrix} \,, \\
    \Delta\mathcal{I}_T & = -\tilde{k_2} \bast \frac{n^2 R^5}{G} \begin{bmatrix} x^2-\frac{1}{2} & xy & xz \\ xy & y^2-\frac{1}{2} & yz \\ xz & yz & 0 \end{bmatrix} \,,
\end{aligned} \end{equation}
where $R$ is the mean radius of Venus, and $\tilde{k_2}$ encompasses the response of Venus's gravitational potential to a perturbation potential. With Eq.~(\ref{eq:products}), we can express their complex representations with only the terms from their third column, $\Delta I = \Delta\mathcal{I}_{xz} + i \Delta\mathcal{I}_{yz}$:
\begin{align} \label{eq:ir_it}
    \Delta I_R & = k_2^{(\mathrm{cw})} \frac{\Omega^2 R^5}{3G} m \,, \\
    \Delta I_T & = k_2^{(\mathrm{cw})} \frac{n^2 R^5}{2G} m + i \frac{n^2 R^5}{2G} \left[ \epsilon \left( k_2^{(1)} e^{i\theta_1} - k_2^{(2)} e^{i\theta_2} \right) - \beta k_2^{(3)} e^{i\theta_3} \right] \,. \nonumber
\end{align}
We find that $\Delta I_R$ is entirely proportional to $m$, whereas $\Delta I_T$ combines a term proportional to $m$ along with three terms with fast forcing frequencies.

\subsubsection{Additional torque on the deformations}
\par The additional solar torque terms that arise from the deformations terms $\Delta\mathcal{I}_R$ and $\Delta\mathcal{I}_T$ are
\begin{equation}
    \vec{\Gamma_R} = 3 \frac{GM_\sun}{\lVert\vec{r}\rVert^5} \vec{r} \times \Delta\mathcal{I}_R \vec{r} \quad \text{and} \quad
    \vec{\Gamma_T} = 3 \frac{GM_\sun}{\lVert\vec{r}\rVert^5} \vec{r} \times \Delta\mathcal{I}_T \vec{r} \,.
\end{equation}
Plugging in the matrices of Eq.~(\ref{eq:ir_it_mat}), with a careful handling of the frequency dependence of the $k_2$ Love number (see Appendix~\ref{app:torques}), we find
\begin{align}
    T_R & = - i k_2^{(\mathrm{cw})} \frac{n^2 \Omega^2 R^5}{2G} m \, - \, \bbar{k_2^{(\mathrm{cw})}} \beta \frac{n^2 \Omega^2 R^5}{2G} e^{i\theta_3} \label{eq:tr} \,, \\
    T_T & = -i \frac{3}{4}\frac{n^4 R^5}{G} \left( k_2^{(\mathrm{cw})}+\bbar{k_2^{(3)}}\!-\!k_2^{(3)} \right) m + \frac{3}{4}\frac{n^4 R^5}{G} \bigg[ - \bbar{k_2^{(\mathrm{cw})}} \beta e^{i\theta_3} \nonumber \\
        & \hspace{3em} + \left( k_2^{(1)}\!+\!\bbar{k_2^{(2)}}\!-\!k_2^{(3)} \right) \epsilon e^{i\theta_1} - \left( \bbar{k_2^{(1)}}\!+\!k_2^{(2)}\!-\!k_2^{(3)} \right) \epsilon e^{i\theta_2} \bigg] \,. \label{eq:tt}
\end{align}

\par The terms proportional to $m$ from $\Delta I_R$, $\Delta I_T$, $T_R$ and $T_R$, from Eqs.~(\ref{eq:ir_it}, \ref{eq:tr}, and \ref{eq:tt}), must be taken into account along with those from Eq.~(\ref{eq:T0}) to fully characterize the free response. This will be achieved in Sect.~\ref{ssec:chandler}.

\subsection{Atmospheric forcing}

\begin{figure}
    \centering
    \includegraphics[width=\hsize]{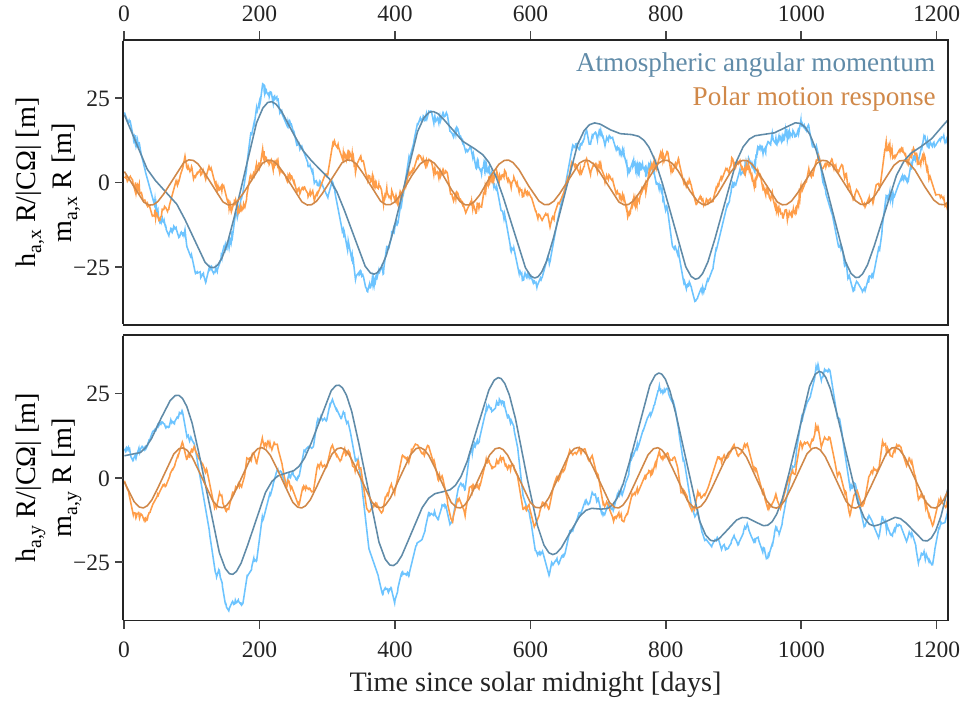}
    \caption{X and Y equatorial components of the atmospheric relative angular momentum ($\Delta h_a$, blue) and of the polar motion it induces on the solid body ($m_a$, orange). The angular momentum is normalized by $|C\Omega|$ which is the transfer function for high frequencies. Bright lines show the numerical time series for the $\Delta h_a$, with $m_a$ computed through a Discrete Fourier Transform. Darker lines show the quasi-periodic approximation for $\Delta h_a$ with only the first three components of Table~\ref{tab:atmosphere_h}, with $m_a$ computed analytically. The time axis is in Earth days.}
    \label{fig:atmosphere}
\end{figure}

\par The atmosphere of Venus affects its solid body spin dynamics through pressure and friction forces on the surface, resulting in a net torque exerted on the solid body. Here, we consider the system composed of the solid planet and the atmosphere, so this forcing is no longer seen as an external torque. Instead, the interaction is controlled by the conservation of total angular momentum. The effect on polar motion appears in Eq.~(\ref{eq:liouville}), where winds vary the angular momentum vector of the atmosphere and mass motions vary its inertia matrix.

\par The Venus Planetary Climate Model (V-PCM) reproduces atmospheric dynamics \citep{lebonnois2010superrotation, lai2024planetary}, by simulating physical, chemical and radiative processes within a 3D grid. Of particular interest here, it computes for each cell the zonal and meridional wind velocities ($u$ and $v$, respectively) along with the density $\rho$, which all contribute to the angular momentum and inertia of the whole atmosphere. The simulation run by \citet{lai2024planetary} provides us with an uninterrupted time series spanning 20~Venus solar days -- more than 6 Earth years. They simulated the dynamics with $240$ time steps per Venus solar day, in a grid of 145 latitudes, 288 longitudes, and 50 vertical levels. Closer to the surface, the levels become progressively thinner and follow the topography, providing a finer vertical resolution in the deep atmosphere.

\par The wind and density quantities contribute to the terms $h$ and $\Delta I$, when spatially integrated over the longitude, $\lambda$, the latitude, $\mu$, and the radius, $r$\,:
\begin{equation} \label{eq:pcm-h} \begin{aligned}
    h_{a,x} & = \int_V r~(u \cos\lambda \sin\mu + v \sin\lambda)~dm \\
    h_{a,y} & = \int_V r~(u \sin\lambda \sin\mu - v \cos\lambda)~dm \\
    h_{a,z} & = \int_V r~(-u \cos\mu)~dm \,,
\end{aligned} \end{equation}
\begin{equation} \label{eq:pcm-i} \begin{aligned}
    \Delta\mathcal{I}_{a,xz} & = \int_V r^2 \cos\lambda \cos\mu \sin\mu ~dm \\
    \Delta\mathcal{I}_{a,yz} & = \int_V r^2 \sin\lambda \cos\mu \sin\mu ~dm \\
    \Delta\mathcal{I}_{a,zz} & = \int_V r^2 \cos^2\!\mu ~dm \,,
\end{aligned} \end{equation}
where $dm \!=\! \rho r^2 \cos\mu~dr~d\mu~d\lambda$, and taking $u$ positive when westward and $v$ positive when northward. This is similar to the approach of \citet{dehant2005atmospheric}, but we integrate vertically using the distance coordinate rather than the pressure coordinate. Here, Eqs.~(\ref{eq:pcm-h},\ref{eq:pcm-i}) use longitudes and latitudes given in the V-PCM coordinate system, attached to the SRF defined in Sect.~\ref{ssec:torque}.

\begin{table}
    \caption{Periodic components of the atmospheric equatorial angular momentum $\Delta h_{a,x} + i \Delta h_{a,y} = \sum_k \Delta h_k \exp(i \nu_k t)$ computed with the V-PCM.} \label{tab:atmosphere_h}
    \centering
    \begin{tabular}{r r r}
    \hline \hline \noalign{\smallskip}
        Period & Amplitude & Phase \\
        $2\pi/\nu_k$~[days] & $|\Delta h_k|$~[$\mathrm{10^{24}~kg~m^2~s^{-1}}$] & $\operatorname{arg}(\Delta h_k)$~[$\degr$] \\
    \noalign{\smallskip} \hline \noalign{\smallskip}
         243.42 & 58.15 &  79 \\
         116.84 & 18.51 & 188 \\
        -116.91 & 12.63 & 259 \\
         307.70 &  4.18 & 199 \\
         208.90 &  4.03 & 286 \\
          58.34 &  3.64 & 187 \\
         180.65 &  3.20 &  12 \\
         398.43 &  2.75 & 193 \\
         -58.57 &  2.71 & 126 \\
         -20.35 &  2.56 &  56 \\
         154.61 &  2.38 & 234 \\
         574.08 &  1.95 & 172 \\
          20.34 &  1.46 & 357 \\
        -389.66 &  1.43 & 248 \\
    \noalign{\smallskip} \hline
    \end{tabular}
    \tablefoot{Time origin for the phases is taken at solar midnight.}
\end{table}

\begin{table}
    \caption{Periodic components of the atmospheric moment of inertia $\Delta I_{a,x} + i \Delta I_{a,y} = \sum_k \Delta I_k \exp(i \nu_k t)$ computed with the V-PCM.} \label{tab:atmosphere_I}
    \centering
    \begin{tabular}{r r r}
    \hline \hline \noalign{\smallskip}
        Period & Amplitude & Phase \\
        $2\pi/\nu_k$~[days] & $|\Delta I_k|$~[$\mathrm{10^{27}~kg~m^2}$] & $\operatorname{arg}(\Delta I_k)$~[$\degr$] \\
    \noalign{\smallskip} \hline \noalign{\smallskip}
         116.60 & 22.89 & 228 \\
        -116.58 & 13.81 &  49 \\
         -14.36 &  7.23 & 175 \\
         -20.32 &  6.58 & 222 \\
         -19.77 &  5.51 & 282 \\
          58.26 &  4.53 &  78 \\
    \noalign{\smallskip} \hline
    \end{tabular}
    \tablefoot{Time origin for the phases is taken at solar midnight.}
\end{table}

\par We obtain time series for $\vec{h_a}$ and $\Delta\mathcal{I}_a$, from which we determine $\vec{h}_{sr}$, and $\vec{\Delta h_a} = \vec{h_a} - \vec{h}_{sr}$. The resulting polar motion response can be computed through the transfer function determined in Sect.~\ref{ssec:forced}, using a Discrete~Fourier~Transform. Additionally, to provide greater interpretability of individual frequency contributions, we approximate each complex-valued time series as a finite sum of periodic components:
\begin{equation}
    \Delta h_a = \sum_k \Delta h_k \, e^{i\nu_k t} \quad \text{and} \quad \Delta I_a = \sum_k \Delta I_k \, e^{i\nu_k t} \,,
\end{equation}
with the process described by \citet{laskar1993frequency}, implemented in the frequency analysis tools provided by TRIP \citep{gastineau2011trip}.

\par Figure~\ref{fig:atmosphere} shows the numerical series for the equatorial angular momentum $\Delta h_a$, along with its analytical approximation consisting of the first three frequency components listed in Table~\ref{tab:atmosphere_h}. The main periodicity is found to be at 243.42 days, closely matching Venus's sidereal rotation period, and the next significant components match the solar day of 116.75 days. The effect of the atmospheric inertia variations $\Delta I_a$ on polar motion is three orders of magnitude smaller that of $\Delta h_a$; the frequencies and amplitudes recovered for $\Delta I_a$ are nonetheless listed in Table~\ref{tab:atmosphere_I}.

\par Excitation functions in the right-hand side of Eq.~(\ref{eq:liouville}) can also be written as sums of periodic components:
\begin{equation} \label{eq:atmosphere_forcing} \begin{aligned}
    i \Delta h_a + \frac{1}{\Omega}\bdot{\Delta h_a} & = i \sum_k \left( 1 + \frac{\nu_k}{\Omega} \right) \Delta h_k \, e^{i\nu_k t} \,, \\
    i \Omega\Delta I_a + \bdot{\Delta I_a} & = i \sum_k \left( \Omega + \nu_k \right) \Delta I_k \, e^{i\nu_k t} \,.
\end{aligned} \end{equation}
These excitation terms, along with those from $T_0$, $\Delta I_T$, $T_R$ and $T_T$ in Eqs.~(\ref{eq:T0},\ref{eq:ir_it},\ref{eq:tr},\ref{eq:tt}), excite the polar motion, whose forced response computation is described in Sect.~\ref{ssec:forced}.

\section{Polar motion solution for Venus}
\label{sec:results}

\par Here, we analytically characterize the period and damping time of Venus's Chandler wobble in Sect.~\ref{ssec:chandler}, as well as the amplitudes and phases of the forced polar motion response in Sect.~\ref{ssec:forced}.

\par The terms not associated with forcing (i.e., those depending on $m$ rather than on forcing frequencies) from the expressions of (i) the solar torque on the mean figure of Venus Eq.~(\ref{eq:T0}); (ii) the centrifugal and tidal deformations Eq.~(\ref{eq:ir_it}); and (iii) the solar torque exerted on these deformations Eqs.~(\ref{eq:tr},\ref{eq:tt}), are all gathered on the left-hand side of Eq.~(\ref{eq:liouville}). The forcing terms are kept on the right-hand side, as a sum of periodic terms. We use the following notations:
\begin{align}
    & s = 1 + \frac{3}{2} \frac{n^2}{\Omega^2} \,, \quad D = k_2^{(\mathrm{cw})} \frac{\Omega^2 R^5}{3G} \,, \\
    & \text{and} \quad D' = -i \imag{k_2^{(3)}} \frac{s-1}{s}\frac{n^2 R^5}{G} \,,
\end{align}
where $s$ is the factor already encountered in Eq.~(\ref{eq:chandler_freq_rigid}) and $D$ is a complex deformation term homogeneous to a moment of inertia having the same phase $\delta_\mathrm{cw}$ as $k_2^{(\mathrm{cw})}$. $D'$ involves the dissipation at the semi-diurnal tide (58 days). We obtain
\begin{equation} \label{eq:liouville_modified} \begin{aligned}
     & \left( \frac{A+B}{2} + sD \right) \bdot{m} - i \Omega s \left(C - \frac{A+B}{2} - sD - D' \right) m \\
     & \hspace{9em} - \frac{B-A}{2} \left( \bdot{\bbar{m}} + i \Omega s \bbar{m} \right) = \sum_k F_k \, e^{i\nu_k t} \,.
\end{aligned} \end{equation}

\subsection{Chandler wobble}
\label{ssec:chandler}

\begin{figure}
    \centering
    \includegraphics[width=\hsize]{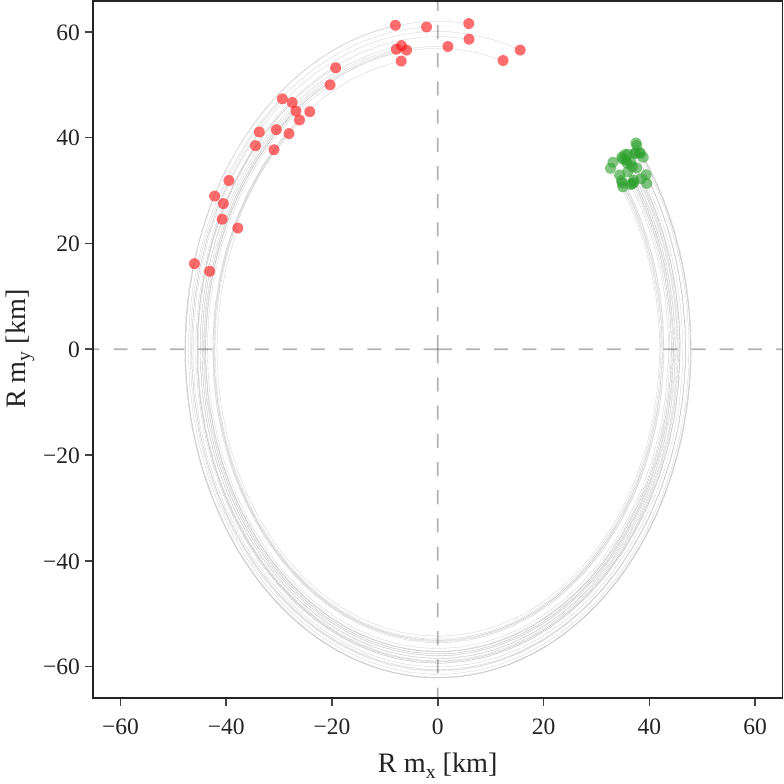}
    \caption{Possible paths of Venus's spin pole in the BRF over 12\,000 years, showing less than one period of the Chandler wobble. The motion corresponds to Eq.~(\ref{eq:chandler_motion_xy}) for a liquid outer core. The thirty green dots show possible initial positions of the pole at epoch J2000. The red dots show the positions after 12\,000 years, their spread reflecting the Chandler period range from Eq.~(\ref{eq:period_result}). Fluctuations due to high-frequency forcing are too small to be seen at this scale (see Fig.~\ref{fig:forced_motion}).}
    \label{fig:chandler_wobble}
\end{figure}

\begin{figure*}
    \centering
    \includegraphics[width=0.6\hsize]{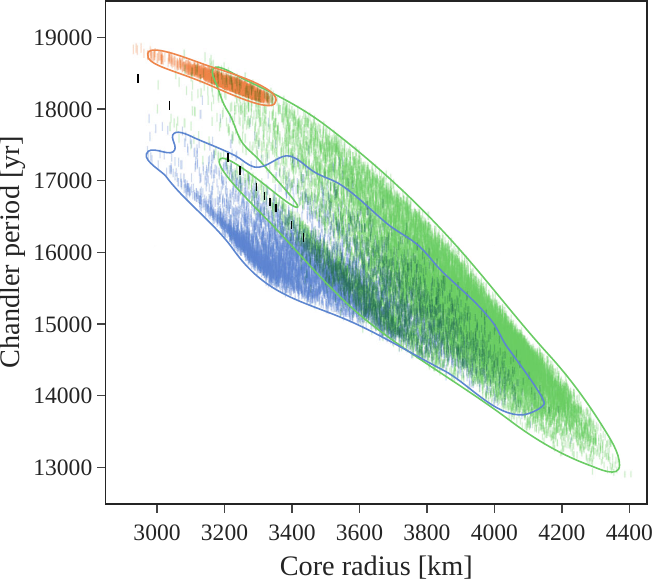}
    \caption{Chandler period of Venus computed for different interior models using Eq.~(\ref{eq:chandler_freq}). Colored bars correspond to the density profiles from \citet{shah2022chemical}, featuring either a fully liquid core (green), a solid inner core with a liquid outer core (blue), or a fully solid core (orange). For each core state, the 99\% contour of a kernel density estimate is shown. Black bars correspond to the density profiles from \citet{dumoulin2017tidal}, all featuring fully liquid cores. The y-axis uncertainties are due to the lengthening of the Chandler period by tidal deformation (from 0.6\% to 1.3\% lengthening).}
    \label{fig:chandler_period}
\end{figure*}

The free solution to Eq.~(\ref{eq:liouville_modified}) has an expression similar to the rigid case solution:
\begin{equation} \label{eq:chandler_motion}
    m(t) = m_0 \exp(i \alpha_\mathrm{cw} t) - \frac{\kappa-1}{\kappa+1} \bbar{m_0} \exp(-i~\bbar{\alpha}_\mathrm{cw} t) \,,
\end{equation}
with the frequency and the ellipticity being
\begin{align} \label{eq:chandler_freq_complex}
    \alpha_\mathrm{cw} & = \Omega s {\left( \frac{C-A-sD-D'}{A+sD} \, \frac{C-B-sD-D'}{B+sD} \right)}^{1/2}, \\[2mm]
    \kappa & \approx \sqrt{\frac{C-A-s\real{D}}{C-B-s\real{D}}} \,.
\end{align}
With the presence of $D$ and $D'$ in $\alpha_\mathrm{cw}$, the frequency now has a complex value that encompasses a damping behavior, as can be seen by writing Eq.~(\ref{eq:chandler_motion}) as
\begin{equation} \label{eq:chandler_motion_xy} \begin{aligned}
    m_x(t) & = m_{x0} \cos(\sigma_\mathrm{cw} t + \eta) \exp(-t/\tau_\mathrm{cw}) \\
    m_y(t) & = \kappa \, m_{x0} \sin(\sigma_\mathrm{cw} t + \eta) \exp(-t/\tau_\mathrm{cw}) \,,
\end{aligned} \end{equation}
with $\alpha_\mathrm{cw} = \sigma_\mathrm{cw} + i/\tau_\mathrm{cw}$. The initial conditions $m_0$, or $(m_{x0}, \eta)$, again must be constrained by observations. Unlike $D$, the term $D'$ has no real part. It thus affects the damping of the free mode, but not the period.

\par Figure~\ref{fig:chandler_wobble} shows different possible solutions for the polar motion of Eq.~(\ref{eq:chandler_motion_xy}). Uncertainties in the wobble period are driven mainly by the uncertainties in the moment of inertia, which we sampled from the interior models of \citet{shah2022chemical} for Fig.~\ref{fig:chandler_wobble}. Uncertainties in the initial conditions $(m_{x0}, \eta)$, in green, are driven by uncertainties of the degree 2 gravitational field, which is taken from \citet{konopliv1999venus}. As recommended in their article, we multiplied the formal uncertainties by a factor of 3, obtaining a distance 50.8km $\pm$ 2.1km between the current spin pole and the figure pole.

\subsubsection{Period of the Chandler wobble}

\par For the wobble frequency, due to the centrifugal and tidal bulges reacting to the wobble, we find an expression that is slightly modified from the rigid case Eq.~(\ref{eq:chandler_freq_rigid}):
\begin{align} \label{eq:chandler_freq}
    \sigma_\mathrm{cw} & = \real{\alpha_\mathrm{cw}} \,, \nonumber \\
    \sigma_\mathrm{cw} & = \Omega \, s \sqrt{\frac{C-A-s\real{D}}{A+s\real{D}} \, \frac{C-B-s\real{D}}{B+s\real{D}}} \,,
\end{align}
where $\real{D} = |D|\cos\delta_\mathrm{cw}$. Again, here, $(A, B, C)$ are either the whole planet's moments of inertia (core, mantle and atmosphere) or those without the core, depending on the core-mantle coupling assumption (see Sect.~\ref{ssec:core}). With our current knowledge of the interior of Venus, the expected Chandler period is not well constrained: Figure~\ref{fig:chandler_period} shows the possible values we obtain for a wide range of interior models.

\par In the case of Venus, the effect of mantle deformation $D$ is small. Using the tidal responses from \citet{musseau2024viscosity} with our Eq.~(\ref{eq:chandler_freq}), we find that the wobble period is slightly increased (compared to the rigid case Eq.~\ref{eq:chandler_freq_rigid}), with increases ranging from 0.6\% for more viscous models to 1.3\% for less viscous models. As noted by \citet{vanhoolst2015rotation}, this effect is minor because the mean shape of Venus is already far from hydrostatic equilibrium, so the tidal and rotational deformations have relatively little effect on the deviation from equilibrium: $s|D|\cos\delta_\mathrm{cw} \ll C-A$.

\par As a result, the values obtained from Eq.~(\ref{eq:chandler_freq}) and from the rigid case Eq.~(\ref{eq:chandler_freq_rigid}) differ by less than 1.5\%. Instead, the main source of uncertainty in determining the Chandler period resides in the value of the moment of inertia. If the outer core is liquid (green or blue in Fig.~\ref{fig:chandler_period}), only the mantle wobbles. Using the mantle moments of inertia from the density profiles of \citet{shah2022chemical}, we find $C_m/(M_VR^2) \in [0.226 ; 0.328]$ when excluding profiles with a fully solid core. This gives a Chandler period range of
\begin{equation} \label{eq:period_result}
    P_\mathrm{cw} = (2\pi/\sigma_\mathrm{cw}) \in [12\,900 ; 18\,800] \text{ years.}
\end{equation}
Otherwise in the case of a fully crystallized core (orange in Fig.~\ref{fig:chandler_period}), the planet wobbles as a whole. Using the moments of inertia from the density profiles of \citet{shah2022chemical}, we find $C/(M_VR^2) \in [0.317 ; 0.329]$ for these solid core cases, yielding a Chandler period in the range $P_\mathrm{cw} \in [18\,100 ; 18\,900]$ years. The ranges for the coupled (fully crystallized core) and decoupled (liquid outer core) cases overlap noticeably, so a measurement of polar motion alone may not be sufficient to distinguish between the two scenarios. This is discussed further in Sect.~\ref{ssec:envision}.

\subsubsection{Damping of the Chandler wobble} \label{sssec:damping}
\par From Eq.~(\ref{eq:chandler_freq_complex}), we derive the characteristic damping time of the wobble. We identify two distinct damping processes, involving the tidal phase lags of both the pole tide and the semi-diurnal tide (58 days), respectively $\delta_\mathrm{cw}$ and $\delta_3$:
\begin{align}
    \dfrac{1}{\tau_\mathrm{cw}} & = \imag{\alpha_\mathrm{cw}} = \dfrac{1}{\tau^{(\mathrm{cw})}} + \dfrac{1}{\tau^{(3)}} \,, \label{eq:chandler_damping}
    \intertext{with}
    \tau^{(\mathrm{cw})} & = \frac{1}{\Omega \, s^2} \frac{f~C}{\imag{-D}} = \frac{1}{\Omega \, s^2} \frac{3GM_V}{\Omega^2 R^3} \frac{f~C}{M_VR^2} \dfrac{1}{-\imag{k_2^{(\mathrm{cw})}}} \,, \label{eq:damping_cw} \\
    \tau^{(3)} & = \frac{1}{\Omega \, s} \frac{f'~C}{\imag{-D'}} = \frac{1}{\Omega \, (s-1)} \frac{GM_V}{n^2 R^3} \frac{f'~C}{M_VR^2} \frac{1}{\imag{k_2^{(3)}}} \,. \label{eq:damping_3}
\end{align}
with $M_V$ and $R$ the mass and radius of Venus, and $f \!=\! AB/C^2$ and $f' \!=\! \frac{AB/C}{(A+B)/2}$ account for the non-sphericity of the planet, and are close to $1$. With an interest only in the order of magnitude for the damping, these factors can be assumed equal to 1.

\par From our phase lag formalism, $\delta_\mathrm{cw}$ and $\Omega$ have opposite signs, which results in $\tau^{(\mathrm{cw})}$ always being positive. Equation~(\ref{eq:damping_cw}) agrees with the expression obtained by \citet{yoder1997venusian}. The effect of the solar torque appears in $\tau^{(\mathrm{cw})}$ through the factor $s^2$, which is around 7.6 for Venus, accounting for almost an order of magnitude faster damping. The tidal dissipation term $\imag{k_2^{(\mathrm{cw})}}$, which should be taken at the period of the Chandler wobble of 16\,000 years, is the least constrained factor in Eq.~(\ref{eq:damping_cw}): the inelastic response of Venus has yet to be measured, and its elastic response is known only at the semi-diurnal period of 58 days. Venus's tidal response can be extrapolated to low frequencies with a range of rheological models (Y. Musseau, priv. comm. 2024), obtaining values spanning nearly two orders of magnitude: $\imag{k_2^{(\mathrm{cw})}} \in [0.003 ; 0.2]$. Including uncertainties in the moment of inertia, we obtain $\tau^{(\mathrm{cw})} \in [0.8~;~80] \times 10^6$ years.

\begin{figure}
    \centering
    \includegraphics[width=\hsize]{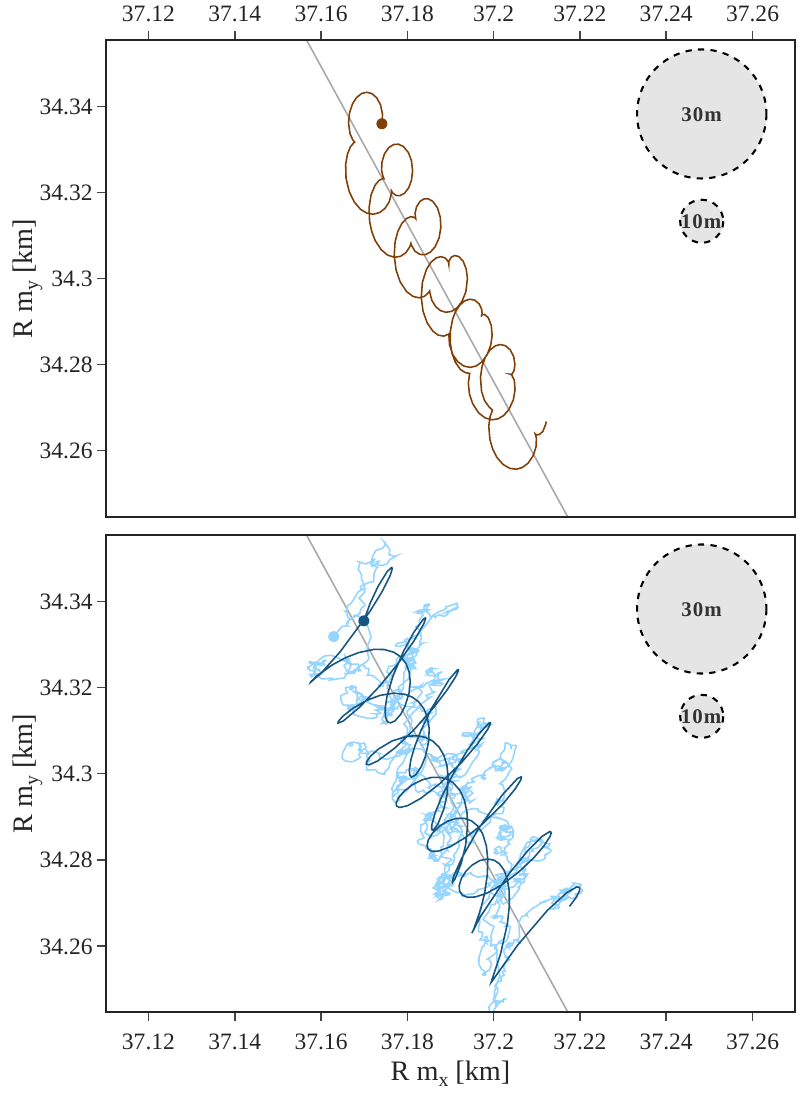}
    \caption{Path of Venus's spin pole in the BRF over 4 years (2034-2038). The gray line represents the free wobble, appearing as a linear trend at this scale. Colored curves show the polar motion caused by: the solar torque only (top panel), the solar torque and the atmosphere (bottom panel, light blue), the solar torque and just the 117-day atmospheric contribution (bottom panel, dark blue, added for readability). Initial positions are indicated with dots. For reference, the surface resolution of EnVision's VenSAR is indicated, for the regional (30m) and targeted (10m) imaging.}
    \label{fig:forced_motion}
\end{figure}

\par Interestingly, for Venus, the least viscous models result in less effective damping (up to 80 Myrs), while the most viscous models lead to more effective damping (down to 0.8 Myr). This can be explained by the fact that, at the Chandler period of around 16\,000 years, for the tidal responses used here, the mantle's behavior begins to approach its fluid limit, where the phase lag $\delta_\mathrm{cw}$ tends toward zero. In this regime, a higher mantle viscosity moves the behavior further from the fluid limit, thereby increasing the phase lag and allowing for a more effective dissipation of the Chandler wobble. For most viscosity profiles, the phase lag of the pole tide is the dominant damping mechanism for Venus, with $\tau^{(\mathrm{cw})} < \tau^{(3)}$.

\par The opposite is true only for the lowest-viscosity end-members where Venus's mantle viscosity is constant at $10^{19}$ Pa\,s, and the phase lag of the semi-diurnal tide becomes the dominant damping mechanism. This additional damping Eq~(\ref{eq:damping_3}) has little effect for fast rotating planets, for which $s$ is close to $1$. The tidal dissipation term $\imag{k_2^{(3)}}$, evaluated at the semi-diurnal period of 58 days, varies by nearly an order of magnitude across the viscosity profiles considered: $-\imag{k_2^{(3)}} \in [0.002 ; 0.02]$, leading to $\tau^{(3)} \in [9.7~;~120] \times 10^6$ years. Combining both contributions in Eq~(\ref{eq:chandler_damping}), with the tidal responses used here (Y. Musseau, priv. comm. 2024), we find an overall Chandler wobble damping time within:
\begin{equation}
    \tau_\mathrm{cw} \in [0.8~;~13] \times 10^6 \text{ years.}
\end{equation}

\par The damping is barely too slow to be noticeable in Fig.~\ref{fig:chandler_wobble}, which covers only 12\,000 years. It is fast enough however, that an excitation process near the Chandler frequency is required to explain the measured angle $0.481\degr \pm 0.020\degr$ \citep{konopliv1999venus} between the spin axis and the figure axis.

\subsection{High-frequency forced polar motion}
\label{ssec:forced}

\begin{table}
    \caption{Periodic components of Venus's forced polar motion response projected on its surface: $R~m_F = \sum_k R~m_k \exp(i \nu_k t)$.} \label{tab:forced_response}
    \centering
    \begin{tabular}{c r r r r}
    \hline \hline \noalign{\smallskip}
        \multirow{2}{*}{Forcing term} & \multirow{2}{*}{Frequency} & Period & \hspace{-1em} Amplitude & Phase \\
        ~ & ~ & [days] & [m] & [$\degr$] \\
    \noalign{\smallskip} \hline \noalign{\smallskip}
        \multirow{3}{*}{$\dfrac{T_0}{\Omega}$}
          & $-\Omega$      &  243.02 & 7.11 & 348 \\
        ~ & $2n-\Omega$    &   76.83 & 2.25 &  56 \\
        ~ & $\Omega$       & -243.02 & 1.80 & 191 \\
    \noalign{\smallskip} \hline \noalign{\smallskip}
        \multirow{6}{*}{$- i \Delta h_a - \dfrac{\bdot{\Delta h_a}}{\Omega}$}
          & $-(n-\Omega)$  & -116.91 & 7.00 & 259 \\
        ~ & $n-\Omega$     &  116.84 & 3.60 & 188 \\
        ~ &                & -389.66 & 1.39 & 248 \\
        ~ & $-2(n-\Omega)$ &  -58.57 & 1.26 & 126 \\
        ~ &                &  -20.35 & 1.04 &  56 \\
        ~ & $2(n-\Omega)$  &   58.34 & 1.04 & 187 \\
    \noalign{\smallskip} \hline
    \end{tabular}
    \tablefoot{Time origin is taken at epoch J2000. Only the components with amplitudes >1m are included.}
\end{table}

\par The polar motion forced response $m_F$ from Eq.~(\ref{eq:liouville_modified}) is
\begin{equation} \label{eq:transfer_function_full} \begin{aligned}
     & m_F(t) = \sum_k F_k e^{i\nu_k t} \, \frac{\Omega s \left(C' - \frac{A'+B'}{2}\right) + \nu_k \frac{A'+B'}{2}}{i~A'B' \left( \nu_k^2 - \alpha_\mathrm{cw}^2 \right)} \\
     & \hspace{10em} + \sum_k \bbar{F} e^{-i\nu_k t} \, \frac{(\Omega s - \nu_k) \frac{B'-A'}{2}}{i~A'B' \left( \nu_k^2 - \alpha_\mathrm{cw}^2 \right)} \,,
\end{aligned} \end{equation}
where $A' \!=\! A\!+\!sD$, \,$B' \!=\! B\!+\!sD$ and $C' \!=\! C\!-\!D'$. Here the complex value of $\alpha_\mathrm{cw}$ again encompasses a damping behavior, preventing an infinite response at $|\nu_k| = |\alpha_\mathrm{cw}|$.

\par The excitation terms found in the solar torque Eqs.~(\ref{eq:T0}, \ref{eq:tr},\ref{eq:tt}) and in the tidal deformation Eq.~(\ref{eq:ir_it}), along with the atmospheric excitation Eq.~(\ref{eq:atmosphere_forcing}), all have periods of a few hundred days maximum. The excitation we consider here is thus in the high-frequency regime when compared to the resonant frequency, with $|\nu_k| \gg |\alpha_\mathrm{cw}| \approx 2\pi/16000$yr. In this regime, considering $(C-A) \ll A$ and $(B-A) \ll A$, the transfer function in Eq.~(\ref{eq:transfer_function_full}) acts as a simple low-pass filter in $1/\nu$:
\begin{align}
     m_F(t) & = -i~\frac{A'+B'}{2A'B'} \sum_k \frac{1}{\nu_k} F_k e^{i\nu_k t} \,, \\
     m_F(t) & \approx -\frac{i}{C} \sum_k \frac{1}{\nu_k} F_k e^{i\nu_k t} \,. \label{eq:transfer_function}
\end{align}
In particular, within $m_F$, the response $m_A$ to the atmospheric excitation terms from Eq.~(\ref{eq:atmosphere_forcing}) can be written as
\begin{equation} \label{eq:transfer_function_atm}
     m_A(t) = -\frac{1}{C} \sum_k \left( \frac{1}{\Omega} + \frac{1}{\nu_k} \right) \left( \Delta h_k + \Omega \Delta I_k \right) e^{i\nu_k t} \,,
\end{equation}
where for higher frequencies $|\nu_k| \gg |\Omega| = 2\pi/243$d, the transfer function becomes constant at $-(C\Omega)^{-1}$. This is convenient when needing to input the raw signal given by the V-PCM, with any high-frequency noise in $\Delta h_a$ simply multiplied by this constant.

\par The transfer function Eq.~(\ref{eq:transfer_function_atm}) cancels out for $\nu = -\Omega$, which happens the be the dominant component of $\Delta h_a$ (Table~\ref{tab:atmosphere_h}). As a consequence, in Fig.~\ref{fig:atmosphere}, the 243-day periodicity in $\Delta h_a$ (blue) is filtered-out in $m_a$ (orange) by the transfer function, leaving the +117 days and -117 days contributions as the dominant components in the atmosphere-induced polar motion. The high-frequency noise is passed unchanged from $\Delta h_a$ to $m_a$, where it appears at the same scale for both in Fig.~\ref{fig:atmosphere} due to the choice of normalization.

\par Given the finite 2333-day span of the atmospheric simulation, small-amplitude components with periods of $\pm$2333 days and $\pm$1167 days appear in $\Delta h_a$, both in the Discrete Fourier Transform and the frequency decomposition. The frequency analysis for such low frequencies is unreliable, and these components could result from a trend or from longer periods generated by the Venus Planetary Climate Model. For this reason, we manually filter out periods longer than 800 days in $\Delta h_a$, as these require longer time series in order to be properly analyzed. If such low frequency truly exist in $\Delta h_a$, they would be significantly amplified in the polar motion through the transfer function Eq.~(\ref{eq:transfer_function_full}).

\par Applying the transfer function Eq.~(\ref{eq:transfer_function}) to each excitation term, we find that the major contributors to the forced polar motion of Venus are the atmospheric winds and the solar torque, as detailed in Table~\ref{tab:forced_response}. Figure~\ref{fig:forced_motion} depicts the polar motion of Venus without and with the atmospheric contribution. When measuring the Chandler frequency, the signal of interest is the average speed of the spin pole along its linear trend (gray line), while the forced fluctuations are regarded as additional noise. The potential detection and measurement of the polar motion by future Venus orbiters is discussed in Sect.~\ref{ssec:envision}.

\par Mantle internal dynamics, core-mantle interactions, and low-frequency atmospheric dynamics could also contribute to the excitation of Venus's polar motion through variations in the mantle's angular momentum and mass distribution. A long-period excitation process is required to explain the current amplitude of its polar motion. It could be a geologically recent event, in which case the observed amplitude would correspond to a purely decaying Chandler wobble, or it could be an ongoing process that actively sustains a large amplitude. Low-frequency excitation is beyond the scope of this work, and further study is needed to determine whether its effects are significant enough to be detected in future measurements.

\section{Discussion}

\subsection{Influence of atmosphere super-rotation}
\label{ssec:superrotation}

\par The wind-induced excess angular momentum, $\vec{h}$, is calculated for Venus using the V-PCM with Eq.~(\ref{eq:pcm-h}). It has a non-zero average, indicating that the atmosphere is in a state of global super-rotation. We find that this average component is close to the spin axis, with an offset of only 127 arcseconds, which led us to assume in Eq.~(\ref{eq:relative_momentum}) that it is aligned with the spin axis. This is a first approach, and a more detailed model of the super-rotation axis orientation could be explored in future work.

\par Here we focus on the effects of super-rotation on polar motion by deriving again Eq.~(\ref{eq:liouville_xyz}) while keeping the terms from the mean super-rotation excess angular momentum, $\vec{h}_{sr}$. We normalize the coefficient $k$ by the mean polar moment of inertia of the atmosphere, $I_a$, introducing a dimensionless super-rotation coefficient $S$ identical to that of \citet{mendonca2016exploring}:
\begin{equation}
    \vec{h}_{sr} = S\;I_a\;\vec{\omega} \,.
\end{equation}
Defined this way, $S$ represents the ratio of the wind-induced angular momentum to the angular momentum of a windless atmosphere. In the absence of super-rotation, $S \!=\! 0$. Using the V-PCM, with the third component of Eqs.~(\ref{eq:pcm-h}) and (\ref{eq:pcm-i}) we compute the mean values $\lVert\vec{h}_{sr}\rVert=2.0\times10^{28}~\mathrm{kg~m^2~s^{-1}}$ and $I_a=1.2\times10^{34}~\mathrm{kg~m^2}$. This corresponds to a global super-rotation coefficient $S=5.7$, which is consistent with the value $S = 7.66^{+4.2}_{-3.6}$ estimated by \citet{mendonca2016exploring} based on available observational data.

\par Equation~(\ref{eq:liouville_xyz}) becomes
\begin{equation} \begin{aligned}
     (A\!+\!SI_a) \bdot{m_x} + \bdot{\Delta\mathcal{I}_{xz}}\!+\!\frac{1}{\Omega} \bdot{\Delta h_x} + \Omega (C\!-\!B) m_y - \Omega \Delta\mathcal{I}_{yz}\!-\!\Delta h_y & \!=\! \frac{1}{\Omega} \Gamma_x \\
     (B\!+\!SI_a) \bdot{m_y} + \bdot{\Delta\mathcal{I}_{yz}}\!+\!\frac{1}{\Omega} \bdot{\Delta h_y} - \Omega (C\!-\!A) m_x + \Omega \Delta\mathcal{I}_{xz}\!+\!\Delta h_x & \!=\! \frac{1}{\Omega} \Gamma_y \\
     (C\!+\!SI_a) \bdot{m_z} + \bdot{\Delta\mathcal{I}_{zz}} + \frac{1}{\Omega} \bdot{\Delta h_z} & \!=\! \frac{1}{\Omega} \Gamma_z \,.
\end{aligned} \end{equation}
Here, we notice that the super-rotation adds $k = SI_a$ to the moments of inertia $(A, B, C)$, essentially increasing the effective inertia of the planet. The Chandler frequency becomes
\begin{equation}
    \sigma_\mathrm{cw} = \Omega \, s \sqrt{\frac{C-A-s\real{D}}{A+s\real{D}+SI_a} \, \frac{C-B-s\real{D}}{B+s\real{D}+SI_a}} \,,
\end{equation}
where the super-rotation slightly increases the Chandler period. We find the effect to be negligible for Venus: it slows down the wobble by a factor of about $S I_a / C_m$, with the super-rotation coefficient $S \!=\! 5.7$ and the atmosphere/mantle inertia ratio of $I_a / C_m \!= 1/4000$, for a overall period increase of only 0.1\%.

\subsection{Influence of core-mantle coupling}
\label{ssec:core}

\par At the core-mantle boundary, a viscosity discontinuity may lead to dynamical decoupling, where the core and the mantle rotate at different speeds and around different axes, exerting a viscous torque on each other. Depending on the core's effective viscosity at the boundary, the Chandler wobble of the mantle may or may not translate to the core. This makes the Chandler period particularly sensitive to the physical state of the core.

\par Previously, we only considered the two limiting cases in which the core of Venus is either decoupled from the mantle (zero torque) or fully locked with it (infinite torque). Here, in this section, we analyze a toy model with a finite coupling coefficient showing that our results in Sect.~\ref{sec:results} hold in both limiting cases, provided that the moments of inertia used in the equations are those of the mantle only in the decoupled case, instead of those of the whole planet that are used in the locked case.

\par Our simplified model in this section is a planet with a rigid axisymmetric mantle and a spherical core whose moment of inertia is $I_c$. The rotation vectors of the mantle and the core, $\vec{\omega}$ and $\vec{\omega_c}$, can differ, resulting in a torque at the core-mantle boundary. A spherical interface implies that the torque exerted on the mantle by the core lies in the direction of the differential rotation \citep{aoki1969friction}:
\begin{equation}
    \vec{\Gamma_c} = k_c (\vec{\omega_c} - \vec{\omega}) \,, \quad \text{with $k_c > 0$,}
\end{equation}
where $k_c$ is the coupling coefficient. As in Sect.~\ref{ssec:torque}, we also take into account the solar torque exerted on the mantle.

\par Similarly to Eq.~(\ref{eq:omega}), the core's spin axis remains close to the z-axis and we write it as $\vec{\omega_c} = \Omega (m_{c,x}, m_{c,y}, 1+m_{c,z})^T$ where $m_{c,x}$, $m_{c,y}$, and $m_{c,z}$ are treated as small quantities. We consider two Euler rotation equations, one for the mantle and one for the whole planet, which become
\begin{equation} \begin{aligned}
    A_m \bdot{m} - i \Omega s (C-A) m & = k_c (m_c - m) \\
    A_m \bdot{m} - i \Omega s (C-A) m & + I_c \bdot{m_c} + i \Omega I_c (m_c - m) = 0 \,,
\end{aligned} \end{equation}
where $A_m = A - I_c$ is the smallest moment of inertia of the mantle. Here, the factor $s = 1 + (3n^2)/(2\Omega^2)$ comes directly from the solar torque Eq.~(\ref{eq:T0}), computed for an axisymmetric mantle. With this system of equations, we find a damped wobble for $m$, whose period and characteristic damping time are
\begin{align}
    \sigma_c & = \Omega s (C - A) \frac{K A + \frac{1}{K} A_m}{K A^2 + \frac{1}{K} A_m^2} \,, \\
    \tau_c & = \frac{1}{\Omega s} \frac{K A^2 + \frac{1}{K} A_m^2}{(C - A) I_c} \,,
\end{align}
where $K = k_c/(I_c \Omega)$ is the normalized coupling coefficient. We note that, while $K$ and $\sigma_c$ have the same sign as $\Omega$, the damping time, $\tau_c$, remains positive.

\begin{figure}
    \centering
    \includegraphics[width=\hsize]{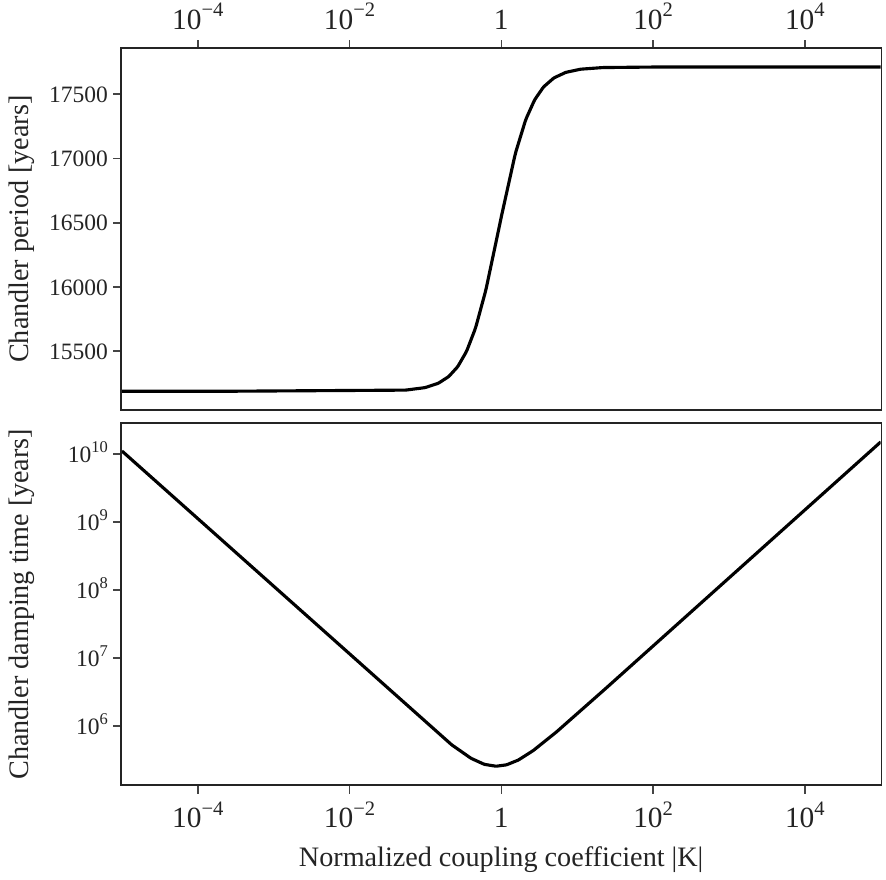}
    \caption{Period and damping time of the Chandler wobble for a simplified Venus (rigid axisymmetric mantle and spherical core), as a function of the core-mantle coupling coefficient.}
    \label{fig:core_coupling}
\end{figure}

\par As expected for the wobble period, in the locked case with $|K| \gg 1$, we find $\sigma_c = \Omega s (C - A)/A$ as for a fully rigid body. In the decoupled case with $|K| \ll 1$, where only the mantle's inertia opposes the wobble, we find a faster wobble $\sigma_c = \Omega s (C - A)/A_m$. Between both regimes, there is a continuous transition at values of $|K|$ around $(A_m/A)^2 \approx 1$, as shown in Fig.~\ref{fig:core_coupling}.

\par It is currently uncertain whether Venus's core is entirely solid, entirely liquid, or consists of a liquid outer core with a solid inner core \citep{dumoulin2017tidal}. If the entire core has crystallized, then $|K| \to \infty$ and Venus wobbles as a single unit. In that case, the Chandler period would serve as a proxy for the total inertia of the planet, similarly to the precession period. On the other hand, if the core has an outer liquid layer or is fully liquid, the coupling coefficient $|K|$ is very weakly constrained. The ranges of kinematic viscosities considered for Venus by both \citet{yoder1997venusian} and \citet{correia2003evolution} indicate that a turbulent friction regime is expected, where $|K| \ll 1$, which implies that a liquid outer core would not wobble along with the mantle. In that case, the Chandler period serves as a proxy for the mantle's inertia.

\par The wobble damping caused by core-mantle friction is most efficient for values of $|K|$ around $(A_m/A)^2 \approx 1$, where it can become the dominant damping process, with the core-mantle damping time $\tau_c$ falling below the $[10^6 ; 10^7]$ years range obtained for the mantle deformation damping time $\tau_\mathrm{cw}$ in Sect.~\ref{sssec:damping}. In the coupled and decoupled limiting cases, however, there is no core-mantle damping occurring, and the dominant process remains mantle deformation damping.

\subsection{Comparison with Earth and Mars}
\label{ssec:earth-mars}

\par Among the planets, the Chandler wobble has been detected on Earth \citep{chandler1891variation} and Mars \citep{konopliv2020detection}. Both planets rotate more than two orders of magnitude faster than Venus and are thus much more oblate, with their $(C-A)/A$ also being more than two orders of magnitude larger than that of Venus. As a result, the expected wobble for Venus is four orders of magnitude slower than the Chandler wobble measured on Earth or Mars (see Table~\ref{tab:earth-mars}). In this slow spin regime, the interior properties that influence a planet's Chandler period differ from those in a fast rotation regime.

\par The solar torque causes a faster wobble, but this effect is significant only for slowly rotating planets (see Sect.~\ref{ssec:torque}). For Earth and Mars, where $\Omega^2 \gg n^2$, the wobble frequency increases by less than 0.001\%, well below current measurement accuracy. In contrast, for Venus, $|\Omega| = 0.92 n$, resulting in a wobble frequency increase by a factor of 2.75.

\par Solid mantle deformations slow down the wobble, lengthening its period by a factor of approximately $|1 - sD/(C-A)|$ (see Sect.~\ref{ssec:deformation}). This effect is more pronounced for fast-rotating planets, where the factor is expressed as $|1 - k_2^{(\mathrm{cw})}/k_2^{(f)}|$, as their oblateness factor $C-A$ is primarily due to the hydrostatic rotational bulge. Here, $k_2^{(f)}$ is the Love number in the fluid limit, at the frequency 0. This leads to a 38\% increase in Earth's Chandler period and a 16\% increase for Mars, when comparing the measured values (Table~\ref{tab:earth-mars}) to those obtained using Eq.~(\ref{eq:chandler_freq_rigid}). For Venus, which rotates more slowly, the non-hydrostatic contribution to $C-A$ becomes dominant, resulting in $|sD/(C-A)| \ll 1$ and reducing the impact of solid deformation. Comparing the values obtained with Eqs. (\ref{eq:chandler_freq_rigid}) and (\ref{eq:chandler_freq}), we find that this effect increases the Chandler period by less than 1.5\%.

\par The amplitude of the Chandler wobble, unlike its period, cannot be solely linked to the planet's internal structure, as it depends on the excitation mechanisms near the Chandler frequency. With an estimate of the damping efficiency of the wobble (see Sect.~\ref{sssec:damping}), measuring the amplitude could help characterize these excitation sources as they sustain the wobble. On Earth, the amplitudes of both free motion (Chandler wobble) and forced motion are comparable, around 150 milliarcseconds \citep{gross2000excitation}. On Mars, the free motion amplitude is 6 milliarcseconds, about four times weaker than the forced motion \citep{konopliv2020detection}. In contrast, Venus's predicted wobble amplitude of 0.5\degr~is four orders of magnitude larger than Earth's. That is three orders of magnitude larger than the forced motion expected from atmospheric dynamics and solar torque.

\par The combined excitation from Earth's ocean and atmosphere near its Chandler period of 433 days has been shown sufficient to sustain its observed wobble \citep{gross2015earth}, essentially through their inertia terms, noted $\Delta I_a$ in our work. For Mars, \citet{dehant2006excitation} identify atmospheric stochastic excitation as the primary excitation source of the wobble at 207 days. For Venus, the source of wobble excitation is still an open question.

\par As a result of the contrasts in the Chandler wobble's periods and amplitudes, the detection methods vary between planets. For the first detection of Earth's Chandler wobble, \citet{chandler1891variation} relied on ground-based stellar observations spanning nearly a full period. For Mars, the wobble was detected through its effect on Mars orbiters trajectories, using spacecraft tracking data spanning 21 years thus covering tens of periods \citep{konopliv2020detection}. For Venus, with an expected Chandler period of around 16\,000 years, even a similar 20-year observational coverage would represent only 0.1\% of a full period. At this timescale, the wobble manifests as a linear trend rather than a periodic signal (see Fig.~\ref{fig:forced_motion}). However, the large amplitude -- 52 km when projected onto Venus's surface -- could still make polar motion detectable by future orbiters, despite its long period.

\begin{table}
    \caption{Comparison of the free polar motion of Venus, Earth, Mars.} \label{tab:earth-mars}
    \centering
    \begin{tabular}{l l r l}
    \hline \hline \noalign{\smallskip}
        ~ & Wobble period & \multicolumn{2}{c}{Wobble amplitude\phantom{------}} \\
        ~ & [years] & [arcsec] & [m] \\
    \noalign{\smallskip} \hline \noalign{\smallskip}
         Venus     & $[12\,900 ; 18\,900]$ & $1790 \pm70$       & $52\,000 \pm2000$ \\
         Earth\tablefootmark{1} & $1.1855 \pm0.0030$    & $0.15 \pm0.05$\tablefootmark{a} & $5 \pm2$\tablefootmark{a}      \\
         Mars\tablefootmark{2}  & $0.5665 \pm 0.0014$   & $0.006 \pm0.001$   & $0.10 \pm0.02$    \\
    \noalign{\smallskip} \hline
    \end{tabular}
    \tablefoot{\tablefoottext{a}{Earth's wobble amplitude is given here with its associated fluctuation. The current measurement uncertainty of Earth's polar motion is 0.1~mas, or 3~mm \citep{bizouard2020geophysical}.}}
    \tablebib{\tablefoottext{1}{\citet{gross2000excitation}}; \tablefoottext{2}{\citet{konopliv2020detection}}}
\end{table}

\subsection{Observations with EnVision}
\label{ssec:envision}

\begin{figure*}
    \centering
    \includegraphics[width=0.53\hsize]{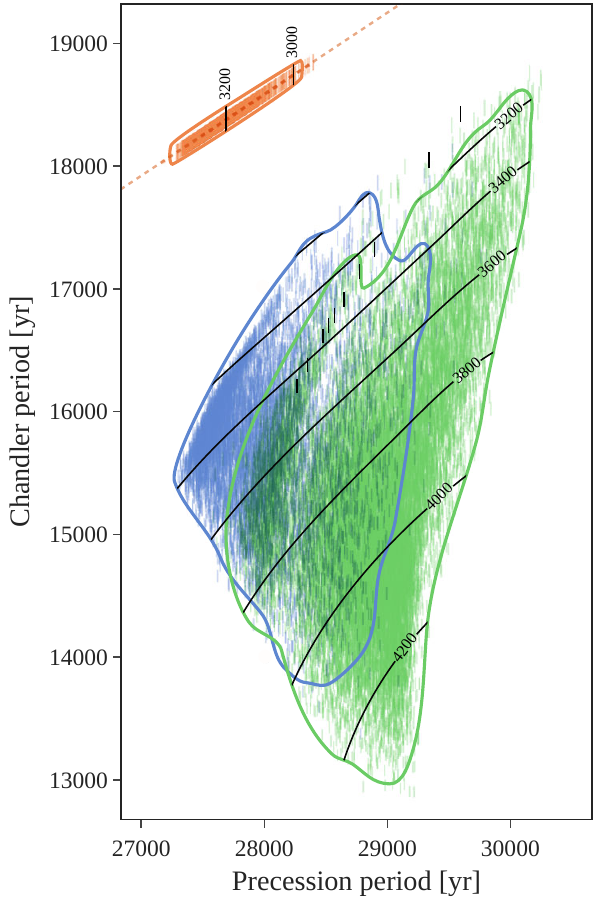}
    \caption{Chandler and precession periods of Venus computed for the density profiles of \citet{shah2022chemical}, featuring either a fully liquid core (green), a solid inner core with a liquid outer core (blue), or a fully solid core (orange). For each core state, the 99\% contour of a kernel density estimate is shown. Black bars correspond to the density profiles from \citet{dumoulin2017tidal}, all featuring fully liquid cores. Black level lines indicate the core size of the models in kilometers, to within 50 km.}
    \label{fig:chandler_precession}
\end{figure*}

\par ESA's EnVision and NASA's VERITAS, two spacecraft currently scheduled for launch no earlier than 2031, are set to study Venus from low, circular, polar orbits, with a focus on geophysics as part of their missions \citep{widemann2023venus}. We focus on EnVision, one of its objectives being to "determine the size of the major internal layers (\dots) and the state of the core, to better understand Venus' thermo-chemical evolution" \citep{esa2023envision}. This will be achieved by its radio science and radar experiments, which will measure Venus's gravity field and rotational dynamics.

\par The synthetic aperture radar, VenSAR, in its imaging mode, will provide repeated views of regions of interest on Venus's surface. This will enable the construction of a geodetic control network by matching radar features across multiple images and solving for their surface coordinates along with a planetary rotation model. For Venus, this approach was first implemented by \citet{davies1992rotation} using Magellan radar images, recovering its rotation period and spin axis orientation, though without sufficient resolution to detect the precession or polar motion signals.

\par The Radio Science Experiment (RSE), in its gravity experiment mode, will produce Doppler tracking data of the EnVision spacecraft, enabling the joint recovery of Venus's gravitational field and spacecraft trajectory. If the gravity field is considered attached to a body-fixed reference frame, a rotation model can be derived simultaneously. Alternatively, if the gravity field is considered attached to the spin reference frame (SRF, see Fig.~\ref{fig:frames}), the polar motion signal is seen instead as variations in the degree-2 gravity coefficients $C_{21}$ and $S_{21}$. For Mars, both approaches were implemented by \citet{konopliv2020detection}, yielding consistent results. For Venus, Magellan tracking data provided a gravity field solution \citep{konopliv1999venus}, albeit only using the predetermined, uniform rotation model of \citet{davies1992rotation}. Recovering Venus's spin axis precession from Doppler tracking data has already been proposed and simulated for the next Venus missions \citep{rosenblatt2021determination, cascioli2021determination}, but the recovery of its polar motion has not been similarly studied.

\par The combined capabilities of VenSAR and RSE could enable a joint inversion of the gravitational field, spacecraft trajectory, control network, and rotation model. \citet{cascioli2021determination} propose this approach for the future VERITAS mission, simulating its VISAR radar and Doppler tracking system, with a rotation model that includes the spin orientation and precession, without the polar motion. They find that the synergy between tie points and Doppler observations should allow for the detection of Venus's precession, with a 0.3\% relative uncertainty for its period.

\par The polar motion model developed here suggests that, over EnVision's 4-year primary mission, Venus's polar motion could be detected alongside its precession. During this period, the spin axis precesses by 8 arcseconds in inertial space, a motion corresponding to 240 meters at the surface, while the Chandler wobble appears as a drift of approximately 90 meters of the spin pole in a body-fixed reference frame. This implies that a surface feature initially coincident with the spin axis, would, after four years, trace circular paths with a radius of 90 meters in inertial space, with a period of 243 days. Such motion should be accounted for: not only to measure the Chandler period, but also to potentially strengthen the determination of the precession period. Additionally, the signal of polar motion could be enhanced by combining future radar and Doppler data from both EnVision and VERITAS. Both Earth-based and Magellan's radar observations have lower resolution, but the long time span between their datasets and those of future missions might still make them valuable for detecting Venus's polar motion.

\par The science operations of EnVision are currently planned to begin in 2034 and could measure both the precession and wobble periods. Figure~\ref{fig:chandler_precession} shows both of these observables for the wide range of interior models produced by \citet{shah2022chemical}. If both periods yield compatible values for Venus's total moment of inertia, it indicates that both motions are opposed by the same total inertia (orange dotted line). It would imply that the core is locked to the mantle even for the Chandler wobble motion, leading to the conclusion that the core is fully solid (see Sect.~\ref{ssec:core}). However, if the two estimates of moments of inertia obtained are incompatible, it would indicate that the core is not fully solid (blue and green regions in Fig.~\ref{fig:chandler_precession}), in which case the moment of inertia derived from the wobble corresponds to the mantle alone. Based on interior modeling \citep[e.g.,][]{shah2022chemical}, measuring both periods would provide a refined estimate of Venus's core size: interior models which have the same total moment of inertia and the same mantle moment of inertia also have similar core sizes, to within 50 km, allowing core size level lines to be drawn in Fig.~\ref{fig:chandler_precession}.

\section{Conclusions}

\par This work provides a comprehensive model of Venus's polar motion. Incorporating the effects of solar torque, mantle viscoelasticity, and atmospheric dynamics, we derived the Chandler wobble period and damping time, as well as the frequencies and amplitudes of the forced motion, with expressions that are valid for both slowly and rapidly rotating terrestrial planets.

\par To apply this model to Venus, we used available measurements \citep{konopliv1999venus, margot2021spin} and relied on interior models \citep{shah2022chemical, musseau2024viscosity}. Our results emphasize the significant role of the solar torque in accelerating the Chandler wobble with a factor of 2.75, predicting a wobble period ranging between 12\,900 and 18\,900 years, depending on the state of Venus's core and the moments of inertia of the mantle and of the whole body. This wobble is damped on timescales ranging from 0.8 to 13 million years, with poor constraints for Venus's dissipation efficiency at these low frequencies.

\par We addressed the high-frequency excitation of Venus's polar motion by atmospheric dynamics and solar torque, which operate at periods up to a few hundred days, predicting the high-frequency polar motion to be accounted for when searching for the free Chandler wobble signal. However, the long-period excitation required to explain the current amplitude should be explored further to determine whether it could contribute to the polar motion measured by future missions.

\par We highlight the necessity of incorporating polar motion into Venus rotation models when anticipating the upcoming orbiter missions. Over EnVision's 4-year mission, the predicted $\sim$ 90-meter polar shift in body-fixed frame due to the Chandler wobble, along with the $\sim$ 240-meter shift in inertial space due to precession, should be measurable. If both motions are indeed detected by the EnVision mission, the physical state of the core would be determined. If Venus's core is not found to be fully solid, the Chandler period is a proxy for the mantle's moment of inertia, providing refined constraints for the size of the core and for thermo-chemical properties of Venus's interior.

\begin{acknowledgements}
We thank R. Helled and O. Shah for providing the Venus density profiles, S. Lebonnois and E. Millour for providing atmospheric simulation data with assistance in reading it, and Y. Musseau for contributing tidal response curves. We thank R.-M. Baland for the careful review.
\end{acknowledgements}

\bibliographystyle{aa}
\bibliography{biblio}

\begin{thebibliography}{40}
\expandafter\ifx\csname natexlab\endcsname\relax\def\natexlab#1{#1}\fi

\bibitem[{Aoki(1969)}]{aoki1969friction}
Aoki, S. 1969, The Astronomical Journal, 74, 284

\bibitem[{Baland {et~al.}(2019)Baland, Coyette, \& {Van
  Hoolst}}]{baland2019coupling}
Baland, R.-M., Coyette, A., \& {Van Hoolst}, T. 2019, Celestial Mechanics and
  Dynamical Astronomy, 131

\bibitem[{Bizouard(2020)}]{bizouard2020geophysical}
Bizouard, C. 2020, Geophysical Modelling of the Polar Motion (De Gruyter)

\bibitem[{Boué {et~al.}(2016)Boué, Correia, \& Laskar}]{boue2016complete}
Boué, G., Correia, A. C.~M., \& Laskar, J. 2016, Celestial Mechanics and
  Dynamical Astronomy, 126, 31

\bibitem[{Cascioli {et~al.}(2021)Cascioli, Hensley, {De Marchi}, Breuer,
  Durante, Racioppa, Iess, Mazarico, \& Smrekar}]{cascioli2021determination}
Cascioli, G., Hensley, S., {De Marchi}, F., {et~al.} 2021, The Planetary
  Science Journal, 2, 220

\bibitem[{Chandler(1891)}]{chandler1891variation}
Chandler, S.~C. 1891, The Astronomical Journal, 11, 59

\bibitem[{Correia {et~al.}(2003)Correia, Laskar, \& {Néron de
  Surgy}}]{correia2003evolution}
Correia, A. C.~M., Laskar, J., \& {Néron de Surgy}, O. 2003, Icarus, 163, 1

\bibitem[{Correia \& Valente(2022)}]{correia2022tidal}
Correia, A. C.~M. \& Valente, E. F.~S. 2022, Celestial Mechanics and Dynamical
  Astronomy, 134

\bibitem[{Cottereau {et~al.}(2011)Cottereau, Rambaux, Lebonnois, \&
  Souchay}]{cottereau2011various}
Cottereau, L., Rambaux, N., Lebonnois, S., \& Souchay, J. 2011, A\&A, 531, A45

\bibitem[{Davies {et~al.}(1992)Davies, Colvin, Rogers, Chodas, Sjogren, Akim,
  Stepanyantz, Vlasova, \& Zakharov}]{davies1992rotation}
Davies, M.~E., Colvin, T.~R., Rogers, P.~G., {et~al.} 1992, JGR Planets, 97,
  13141

\bibitem[{Dehant {et~al.}(2005)Dehant, {de Viron}, \&
  {Greff-Lefftz}}]{dehant2005atmospheric}
Dehant, V., {de Viron}, O., \& {Greff-Lefftz}, M. 2005, A\&A, 438, 1149

\bibitem[{Dehant {et~al.}(2006)Dehant, {de Viron}, Karatekin, \& {Van
  Hoolst}}]{dehant2006excitation}
Dehant, V., {de Viron}, O., Karatekin, O., \& {Van Hoolst}, T. 2006, A\&A, 446,
  345

\bibitem[{Dumoulin {et~al.}(2017)Dumoulin, Tobie, Verhoeven, Rosenblatt, \&
  Rambaux}]{dumoulin2017tidal}
Dumoulin, C., Tobie, G., Verhoeven, O., Rosenblatt, P., \& Rambaux, N. 2017,
  JGR Planets, 122, 1338

\bibitem[{Efroimsky(2012)}]{efroimsky2012bodily}
Efroimsky, M. 2012, Celestial Mechanics and Dynamical Astronomy, 112, 283

\bibitem[{ESA(2023)}]{esa2023envision}
ESA. 2023, EnVision - Understanding why Earth's closest neighbour is so
  different, Definition Study Report (Red Book), ESA document reference

\bibitem[{Gastineau \& Laskar(2011)}]{gastineau2011trip}
Gastineau, M. \& Laskar, J. 2011, ACM Communications in Computer Algebra, 44,
  194

\bibitem[{Gross(2000)}]{gross2000excitation}
Gross, R.~S. 2000, Geophysical Research Letters, 27, 2329

\bibitem[{Gross(2015)}]{gross2015earth}
Gross, R.~S. 2015, in Treatise on Geophysics (Elsevier), 215--261

\bibitem[{Konopliv {et~al.}(1999)Konopliv, Banerdt, \&
  Sjogren}]{konopliv1999venus}
Konopliv, A.~S., Banerdt, W.~B., \& Sjogren, W.~L. 1999, Icarus, 139, 3

\bibitem[{Konopliv {et~al.}(2020)Konopliv, Park, Rivoldini, Baland, {Le
  Maistre}, {Van Hoolst}, Yseboodt, \& Dehant}]{konopliv2020detection}
Konopliv, A.~S., Park, R.~S., Rivoldini, A., {et~al.} 2020, Geophysical
  Research Letters, 47

\bibitem[{Konopliv \& Yoder(1996)}]{konopliv1996venusian}
Konopliv, A.~S. \& Yoder, C.~F. 1996, Geophysical Research Letters, 23, 1857

\bibitem[{Lai {et~al.}(2024)Lai, Lebonnois, \& Li}]{lai2024planetary}
Lai, D., Lebonnois, S., \& Li, T. 2024, JGR Planets, 129

\bibitem[{Laskar(1993)}]{laskar1993frequency}
Laskar, J. 1993, Celestial Mechanics and Dynamical Astronomy, 56, 191

\bibitem[{Lebonnois {et~al.}(2010)Lebonnois, Hourdin, Eymet, Crespin, Fournier,
  \& Forget}]{lebonnois2010superrotation}
Lebonnois, S., Hourdin, F., Eymet, V., {et~al.} 2010, JGR Planets, 115

\bibitem[{Margot {et~al.}(2021)Margot, Campbell, Giorgini, Jao, Snedeker,
  Ghigo, \& Bonsall}]{margot2021spin}
Margot, J.-L., Campbell, D.~B., Giorgini, J.~D., {et~al.} 2021, Nature
  Astronomy, 5, 676

\bibitem[{Mendonça \& Read(2016)}]{mendonca2016exploring}
Mendonça, J.~M. \& Read, P.~L. 2016, Planetary and Space Science, 134, 1

\bibitem[{Munk \& MacDonald(1960)}]{munk1960rotation}
Munk, W.~H. \& MacDonald, G.~J. 1960, The Rotation of the Earth: a geophysical
  discussion (Cambridge University Press)

\bibitem[{Musseau {et~al.}(2024)Musseau, Tobie, Dumoulin, Gillmann, Revol, \&
  Bolmont}]{musseau2024viscosity}
Musseau, Y., Tobie, G., Dumoulin, C., {et~al.} 2024, Icarus, 116245

\bibitem[{Rambaux(2013)}]{rambaux2013rotational}
Rambaux, N. 2013, A\&A, 556, A151

\bibitem[{Rambaux {et~al.}(2011)Rambaux, {Castillo-Rogez}, Dehant, \&
  Kuchynka}]{rambaux2011constraining}
Rambaux, N., {Castillo-Rogez}, J., Dehant, V., \& Kuchynka, P. 2011, A\&A, 535,
  A43

\bibitem[{Rosenblatt {et~al.}(2021)Rosenblatt, Dumoulin, Marty, \&
  Genova}]{rosenblatt2021determination}
Rosenblatt, P., Dumoulin, C., Marty, J.-C., \& Genova, A. 2021, Remote Sensing,
  13, 1624

\bibitem[{Saliby {et~al.}(2023)Saliby, Fienga, Briaud, Mémin, \&
  Herrera}]{saliby2023viscosity}
Saliby, C., Fienga, A., Briaud, A., Mémin, A., \& Herrera, C. 2023, Planetary
  and Space Science, 231, 105677

\bibitem[{Shah {et~al.}(2022)Shah, Helled, Alibert, \&
  Mezger}]{shah2022chemical}
Shah, O., Helled, R., Alibert, Y., \& Mezger, K. 2022, The Astrophysical
  Journal, 926, 217

\bibitem[{Spada {et~al.}(1996)Spada, Sabadini, \& Boschi}]{spada1996spin}
Spada, G., Sabadini, R., \& Boschi, E. 1996, Geophysical Research Letters, 23,
  1997

\bibitem[{{Van Hoolst}(2015)}]{vanhoolst2015rotation}
{Van Hoolst}, T. 2015, in Treatise on Geophysics (Elsevier), 121--151

\bibitem[{Widemann {et~al.}(2023)Widemann, Smrekar, Garvin, {Straume-Lindner},
  Ocampo, Schulte, Voirin, Hensley, Dyar, Whitten, Nunes, Getty, Arney,
  Johnson, Kohler, Spohn, {O'Rourke}, Wilson, Way, Ostberg, Westall, Höning,
  Jacobson, Salvador, Avice, Breuer, Carter, Gilmore, Ghail, Helbert, Byrne,
  Santos, Herrick, Izenberg, Marcq, Rolf, Weller, Gillmann, Korablev, Zelenyi,
  Zasova, Gorinov, Seth, Rao, \& Desai}]{widemann2023venus}
Widemann, T., Smrekar, S.~E., Garvin, J.~B., {et~al.} 2023, Space Science
  Reviews, 219

\bibitem[{Williams(1994)}]{williams1994contributions}
Williams, J.~G. 1994, The Astronomical Journal, 108, 711

\bibitem[{Williams {et~al.}(2001)Williams, Boggs, Yoder, Ratcliff, \&
  Dickey}]{williams2001lunar}
Williams, J.~G., Boggs, D.~H., Yoder, C.~F., Ratcliff, J.~T., \& Dickey, J.~O.
  2001, JGR Planets, 106, 27933

\bibitem[{Yoder(1997)}]{yoder1997venusian}
Yoder, C.~F. 1997, in Venus II (University of Arizona Press), 1087--1124

\bibitem[{Yoder \& Ward(1979)}]{yoder1979venus}
Yoder, C.~F. \& Ward, W.~R. 1979, The Astrophysical Journal, 233, L33

\end{thebibliography}

\onecolumn
\begin{appendix}

\section{Numerical parameters used for Venus}

\begin{table}[!ht]
    \caption{Numerical parameters used for Venus} \label{tab:venus_numbers}
    \centering
    \begin{tabular}{r l l l}
    \hline \hline \noalign{\smallskip}
        Parameter & ~ & Value & Reference \\
    \noalign{\smallskip} \hline \noalign{\smallskip}
Orbital period               & $2\pi/n$                     & $224.7$ days                                 & ~    \\
Sidereal rotation period     & $2\pi/\Omega$                & $243.0226(13)$ days                          & \tablefootmark{1} \\
Gravitational parameter      & $GM_V$                       & $3.24858592(6) \times 10^{14}$ m$^3$s$^{-2}$ & \tablefootmark{2} \\
$J_2$\tablefootmark{a}       & $(C-\frac{B+A}{2})/(M_VR^2)$ & $4.4049(45) \times 10^{-6}$                  & \tablefootmark{2} \\
$C_{22}$\tablefootmark{a}    & $(B-A)/(4M_VR^2)$            & $5.571(19) \times 10^{-7}$                   & \tablefootmark{2} \\
Normalized moment of inertia & $C/(M_VR^2)$                 & $0.337(24)$                                  & \tablefootmark{1} \\
Obliquity                    & $\epsilon$                   & $2.6392(8)\degr$                             & \tablefootmark{1} \\
Init. spin-axis offset       & $\beta_{J2000}$              & $0.481(20)\degr$                             & \tablefootmark{2} \\
\multirow{2}{*}{Init. spin-axis direction \bigg\{}
                             & $\alpha_{J2000}$             & $46.5\degr(2.4\degr)$                        & \tablefootmark{2} \\
 ~                           & $\gamma_{J2000}$             & $-49.7\degr(2.4\degr)$                       & \tablefootmark{2} \\
    \noalign{\smallskip} \hline
    \end{tabular}
    \tablefoot{\tablefoottext{a}{$J_2$ and $C_{22}$ here are expressed in the BRF where the inertia matrix is diagonal, thus are slightly higher than in \citet{konopliv1999venus}, with relative differences of $10^{-4}$ and $6\times10^{-3}$ respectively.}}
    \tablebib{\tablefoottext{1}{\citet{margot2021spin}}; \tablefoottext{2}{\citet{konopliv1999venus}}}
\end{table}

\section{Derivation of the torque acting on the tidal bulges}
\label{app:torques}

\par Here, we use the notation $\tidal{X}$ to denote the tidal response of Venus to a perturbation $X$:
\begin{equation}
    \tidal{X}(t) = \tilde{k_2}(\tau) \bast X(t) = \int_0^{+\infty} \! \tilde{k_2}(\tau) X(t-\tau) d\tau \,,
\end{equation}
where $\tilde{k_2}(\tau)$ is the convolution kernel that encompasses the body's rheological properties. With this notation, perturbations with opposite frequencies are associated with conjugate love numbers \citep[e.g.,][]{efroimsky2012bodily}: if $\tidal{e^{i\nu_j t}} = k_2^{(j)}e^{i\nu_j t}$ then $\tidal{e^{-i\nu_j t}} = \bbar{k_2^{(j)}}e^{-i\nu_j t}$. With conjugate love numbers having opposite phase lags, then all time lags remain negative, as physically expected.

\par For clarity, the tidal deformation is first expressed with the notation $\{.\}$, and we switch to using the individual $k_2^{(j)}$ values from Table~\ref{tab:frequencies} when the frequencies are separately identified.

\subsection{Solar torque on the centrifugal bulge}
Using the coordinates $\vec{r}$ of the Sun in the BRF (Eq.~\ref{eq:sun_xyz}), we derive the solar torque $\vec{\Gamma_R}$ acting on Venus's centrifugal deformation:
\begin{align}
    \vec{\Gamma_R} &= 3 \frac{GM_\sun}{\lVert\vec{r}\rVert^5} \vec{r} \times \Delta\mathcal{I}_R \vec{r} \,, \quad \text{with} \quad \Delta\mathcal{I}_R = \frac{\Omega^2 R^5}{3G} \begin{bmatrix} \tidal{-\frac{2}{3}m_z} & 0 & \tidal{m_x} \\ 0 & \tidal{-\frac{2}{3}m_z} & \tidal{m_y} \\ \tidal{m_x} & \tidal{m_y} & \tidal{\frac{4}{3}m_z} \end{bmatrix} \,, \quad \text{and} \quad \frac{\vec{r}}{\lVert\vec{r}\rVert} = \begin{bmatrix} x \\ y \\ z \end{bmatrix} \,, \\
    \vec{\Gamma_R} &= \frac{n^2\Omega^2 R^5}{G} \begin{bmatrix} xy\tidal{m_x} + y^2\tidal{m_y} \\ -x^2\tidal{m_x} - xy\tidal{m_y} \\ 0 \end{bmatrix} \,,
    \intertext{where we considered only the first-order terms in $\epsilon$ and $\beta$. With $T_R = \Gamma_{R,x} + i \Gamma_{R,y}$, we have}
    T_R &= \frac{n^2\Omega^2 R^5}{2G} \left[ (y^2 + x^2)\tidal{m_y - i m_x} + (y^2 - x^2 - 2 i xy)\tidal{m_y + i m_x} \right] \,, \\
    T_R &= \frac{n^2\Omega^2 R^5}{2G} \left[\tidal{\beta e^{-i\alpha}} - e^{i(\theta_3-\alpha)}\tidal{\beta e^{i\alpha}} \right] \,.
    \intertext{We then finally obtain}
    T_R &= -i k_2^{(\mathrm{cw})} \frac{n^2\Omega^2 R^5}{2G} m - \bbar{k_2^{(\mathrm{cw})}} \beta \frac{n^2\Omega^2 R^5}{2G} e^{i\theta_3} \,.
\end{align}

\subsection{Solar torque on the solar tidal bulge}
We then derive the solar torque $\vec{\Gamma_T}$ acting the solar tidal deformation:
\begin{align}
    \vec{\Gamma_T} &= 3 \frac{GM_\sun}{\lVert\vec{r}\rVert^5} \vec{r} \times \Delta\mathcal{I}_T \vec{r} \,, \quad \text{with} \quad \Delta\mathcal{I}_T = -\frac{n^2 R^5}{G} \begin{bmatrix}
        \tidal{x^2-\frac{1}{2}} & \tidal{xy} & \tidal{xz} \\
        \tidal{xy} & \tidal{y^2-\frac{1}{2}} & \tidal{yz} \\
        \tidal{xz} & \tidal{yz} & 0
    \end{bmatrix} \,, \quad \text{and} \quad \frac{\vec{r}}{\lVert\vec{r}\rVert} = \begin{bmatrix} x \\ y \\ z \end{bmatrix} \,. \\
    \vec{\Gamma_T} &= -3\frac{n^4 R^5}{G} \begin{bmatrix}
        xy\tidal{xz} + y^2\tidal{yz} - xz\tidal{xy} - yz\tidal{y^2-\frac{1}{2}} \\
        xz\tidal{x^2-\frac{1}{2}} + yz\tidal{xy} - x^2\tidal{xz} - xy\tidal{yz} \\
        x^2\tidal{xy} + xy\tidal{y^2-\frac{1}{2}} - xy\tidal{x^2-\frac{1}{2}} - y^2\tidal{xy} \\
    \end{bmatrix} \,,
    \intertext{where we considered only the first-order terms in $\epsilon$ and $\beta$. With $T_T = \Gamma_{T,x} + i \Gamma_{T,y}$, we have}
    T_T &= -\frac{3}{2}\frac{n^4 R^5}{G} \left[ (y^2+x^2)\tidal{yz-ixz} - (x^2-y^2+2ixy)\tidal{yz+ixz} + (yz+ixz)\tidal{x^2-y^2+2ixy} - (yz-ixz)\tidal{y^2+x^2-1} \right] \,, \\
    T_T &= -\frac{3}{4}\frac{n^4 R^5}{G} \bigg[ \tidal{-\epsilon e^{i\theta_1} + \epsilon e^{i\theta_2} + \beta e^{i\theta_3} - \beta e^{-i\alpha}} \nonumber \\
        & \hspace{9em} - e^{i(\theta_3-\alpha)}\tidal{-\epsilon e^{-i\theta_1} + \epsilon e^{-i\theta_2} + \beta e^{-i\theta_3} - \beta e^{i\alpha}} + (-\epsilon e^{-i\theta_1} + \epsilon e^{-i\theta_2} + \beta e^{-i\theta_3} - \beta e^{i\alpha})\tidal{e^{i(\theta_3-\alpha)}} \bigg] \,.
    \intertext{Noticing that $\theta_1+\theta_2 = \theta_3-\alpha$,}
    T_T &= -\frac{3}{4}\frac{n^4 R^5}{G} \bigg[ \left( -k_2^{(1)}\epsilon e^{i\theta_1} + k_2^{(2)}\epsilon e^{i\theta_2} + k_2^{(3)}\beta e^{i\theta_3} - k_2^{(\mathrm{cw})}\beta e^{-i\alpha} \right) \nonumber \\
        & \hspace{9em} - \left( -\bbar{k_2^{(1)}}\epsilon e^{i\theta_2} + \bbar{k_2^{(2)}}\epsilon e^{i\theta_1} + \bbar{k_2^{(3)}}\beta e^{-i\alpha} - \bbar{k_2^{(\mathrm{cw})}}\beta e^{i\theta_3} \right) + \left( -k_2^{(3)}\epsilon e^{i\theta_2} + k_2^{(3)}\epsilon e^{i\theta_1} + k_2^{(3)}\beta e^{-i\alpha} - k_2^{(3)}\beta e^{i\theta_3} \right) \bigg] \,.
    \intertext{We then finally obtain}
    T_T &= -i \frac{3}{4}\frac{n^4 R^5}{G} \left( k_2^{(\mathrm{cw})}+\bbar{k_2^{(3)}}-k_2^{(3)} \right) m + \frac{3}{4}\frac{n^4 R^5}{G} \left[ \left( k_2^{(1)}+\bbar{k_2^{(2)}}-k_2^{(3)} \right) \epsilon e^{i\theta_1} - \left( \bbar{k_2^{(1)}}+k_2^{(2)}-k_2^{(3)} \right) \epsilon e^{i\theta_2} - \bbar{k_2^{(\mathrm{cw})}} \beta e^{i\theta_3} \right] \,.
\end{align}

\end{appendix}
\twocolumn

\end{document}